\documentclass[aps,prl,twocolumn,superscriptaddress]{revtex4-2}

\usepackage[utf8]{inputenc} % allow utf-8 input
\usepackage[T1]{fontenc}    % use 8-bit T1 fonts
\usepackage{hyperref}       % hyperlinks
\usepackage{url}            % simple URL typesetting
\usepackage{booktabs}       % professional-quality tables
\usepackage{amsfonts}       % blackboard math symbols
\usepackage{amsmath}
\usepackage{amsthm}
\usepackage{amssymb}
\usepackage{mathtools}
\usepackage{commath}
\usepackage{mathrsfs}
\usepackage{bbm}
\usepackage{nicefrac}       % compact symbols for 1/2, etc.
\usepackage{microtype}      % microtypography
\usepackage{lipsum}
\usepackage{graphicx}
\graphicspath{ {./images/} }
\usepackage{xcolor}
\usepackage[toc]{appendix}
\usepackage{enumitem}   
\usepackage{comment}
\usepackage{footmisc}

\usepackage{fancyhdr}
\fancypagestyle{firstpagefooter}{
    \fancyhf{} % Clear all header and footer fields
    \fancyfoot[C]{Submitted to Physical Review Letters} % Centered footer text
}

\newlist{primenumerate}{enumerate}{1}
\setlist[primenumerate,1]{label={\arabic*$'.$}}
\usepackage{bibunits}
\newtheorem{theorem}{Theorem}
\newtheorem{corollary}{Corollary}
\newtheorem{lemma}{Lemma}
\newtheorem{definition}{Definition}
\newcommand{\J}{\mathbf{J}}
\newcommand{\R}{\mathbb{R}}
\newcommand{\E}{\mathbb{E}}
\newcommand{\Var}{\text{Var}}
\newcommand{\by}{\mathbf{y}}
\newcommand{\bz}{\mathbf{z}}
\newcommand{\bxi}{\boldsymbol{\xi}}

\newcommand{\rd}{\mathrm{d}}

\begin{document}

%\preprint{}

%\title{Convergence of duplicate-free spiking neural networks to rate networks}
%\title{Convergence of duplicate-free spiking neural networks to rate networks}
%\title{Convergence of duplicate-free spiking neural networks to rate-based dynamics}
\title{Emergent rate-based dynamics in duplicate-free populations of spiking neurons}
%\title{From spike-based interaction to rate-based population dynamics: concentration of measure in duplicate-free networks}
\author{Valentin Schmutz}
\email[Corresponding author: ]{v.schmutz@ucl.ac.uk} 
\affiliation{School of Life Sciences and School of Computer and Communication Sciences, École Polytechnique Fédérale de Lausanne, 1015 Lausanne, Switzerland}
\affiliation{UCL Queen Square Institute of Neurology, University College London, WC1E 6BT London, United Kingdom}

\author{Johanni Brea}
\author{Wulfram Gerstner}
%\author{Valentin Schmutz\footnote{Corresponding author: v.schmutz@ucl.ac.uk}, Johanni Brea, Wulfram Gerstner\footnote{Corresponding author: wulfram.gerstner@epfl.ch}}
\affiliation{School of Life Sciences and School of Computer and Communication Sciences, École Polytechnique Fédérale de Lausanne, 1015 Lausanne, Switzerland}

\begin{abstract}
{\color{black}Can Spiking Neural Networks} (SNNs) approximate the dynamics of Recurrent Neural Networks (RNNs)? Arguments in classical mean-field theory based on laws of large numbers provide a positive answer when each neuron in the network has many ``duplicates'', i.e. other neurons with almost perfectly correlated inputs. Using a disordered network model that guarantees the absence of duplicates, we show that duplicate-free SNNs can converge to RNNs, thanks to the concentration of measure phenomenon. {\color{black}This result reveals a general mechanism underlying the emergence of rate-based dynamics in large SNNs.}
%Can the dynamics of Spiking Neural Networks (SNNs) approximate the dynamics of Recurrent Neural Networks (RNNs)? Arguments in classical mean-field theory based on laws of large numbers provide a positive answer when each neuron in the network has many ``duplicates'', i.e. other neurons with almost perfectly correlated inputs. Using a disordered network model that guarantees the absence of duplicates, we show that duplicate-free SNNs can converge to RNNs, thanks to the concentration of measure phenomenon. This result broadens the mechanistic plausibility of rate-based theories in neuroscience. 
%Can the dynamics of large Spiking Neural Networks (SNNs) approximate the dynamics of equally large Recurrent Neural Networks (RNNs)? Arguments in classical mean-field theory based on laws of large numbers provide a positive answer, when each neuron in the network has many ``duplicates'', i.e. other neurons with almost perfectly correlated inputs. Using a disordered network model which guarantees the absence of neuronal duplicates in large networks, we show that the dynamics of duplicate-free SNNs can still converge to that of RNNs, thanks to the concentration of measure phenomenon.

\end{abstract}

\maketitle
\thispagestyle{firstpagefooter}
%\begin{bibunit}
Neurons in the brain interact via spikes -- short and stereotyped membrane potential deflections -- commonly modelled as Dirac pulses \cite{RieWar97, GerKis02}. SNNs with recurrent connectivity are simplified models of real networks with the essential biological feature of spike-based neuronal communication. {\color{black}In contrast}, traditional RNNs are continuous dynamical systems {\color{black}in which} abstract rate neurons directly transmit their firing rate to other neurons, a type of communication which is not biological. Despite their inferior realism, RNNs continue to play a central role in theoretical neuroscience because they can be trained by modern machine learning methods \cite{Bar17}, they can be analysed using tools from statistical physics {\color{black}\cite{AmiGut85b,SomCri88,KuhBos91,MasOst18,PerAlj23}}, and because biological networks are believed to implement computations by approximation of continuous dynamical systems \cite{VyaGol20}. Closing the gap between the more biological SNNs and the more tractable RNNs {\color{black}has remained a challenging theoretical problem \cite{Bre15}}. {\color{black}This problem is important for neuroscience as it is linked to a fundamental question: given the discontinuous and stochastic nature of spike-based communication \cite{FaiSel08}, how can biological networks produce noise-robust population dynamics \cite{LonRot10, MonWol12}?}

%In this letter, we use a transparent modeling framework to show that weak connectivity between neurons is sufficient for a SNN to approximate a RNN.

To state the problem, let us consider an SNN composed of $N$ {\color{black}linear-nonlinear-Poisson neurons} {\color{black}\cite{BreMas96,Chi01}}. For each neuron index $i$, the spike times $\{t^k_i\}_k$ of neuron $i$, which define the neuron's spike train $S_i(t) = \sum_k \delta(t-t^k_i)$, are generated by an inhomogeneous Poisson process with instantaneous firing rate $\phi(h_i(t))$, where $h_i(t)$ represents the neuron's potential and $\phi$ is a positive-valued nonlinear transfer function. The potential $h_i(t)$ is a leaky integrator of the recurrent inputs coming from neurons $j \neq i$ and the external input $I^{\mathrm{ext}}_i(t)$:
\begin{equation}\label{eq:SNN}
    \tau \frac{\rd}{\rd t}h_i(t) = - h_i(t) + \sum_{j=1}^N J_{ij}S_j(t)+ I^{\mathrm{ext}}_i(t),
\end{equation}
where $\tau$ is the integration (or membrane) time constant and $J_{ij}$ is the synaptic weight from neuron~$j$ to neuron~$i$ (by convention, $J_{ii}=0$). While the spike-based model described here is biologically simplistic, it is mathematically convenient as it has a straightforward rate-based counterpart. If we replace the spike trains $\{S_j(t)\}_j$ in~\eqref{eq:SNN} by the corresponding instantaneous firing rates $\{\phi(h_j(t))\}_j$, we get the rate-based dynamics
\begin{equation}\label{eq:RNN}
    \tau \frac{\rd}{\rd t}x_i(t) = - x_i(t) + \sum_{j=1}^N J_{ij}\phi(x_j(t)) + I^{\mathrm{ext}}_i(t),
\end{equation}
which defines an RNN with $N$ rate units. To avoid confusion, we write $h_i(t)$ for the potentials of the SNN~\eqref{eq:SNN} and $x_i(t)$ for the potentials of the RNN~\eqref{eq:RNN}. While the mapping from the SNN to the RNN looks simple at first glance, the spike-based stochastic process~\eqref{eq:SNN} and the rate-based dynamical system~\eqref{eq:RNN} describe very different kinds of systems and the SNN potentials $h_i(t)$ are not guaranteed, in general, to be equal or even close to the RNN potentials $x_i(t)$. 

There are two known types of scaling limits where the SNN potentials $h_i(t)$ converge to the RNN potentials $x_i(t)$:
\begin{enumerate}[label=(\roman*)]
    \item \textit{Spatial averaging over neuronal duplicates}: Consider networks of increasing size $N$ where neurons are localized in some fixed space such that two neurons assigned to the same point are duplicates, i.e. they always share the same recurrent and external input. If the synaptic weights are scaled by $1/N$, we can take the mean-field limit $N\to\infty$ \cite{Ger95}. The fixed space can be either discrete and finite \cite{DitLoe17} or continuous and finite-dimensional \cite{CheDua19}, e.g. a ring \cite{BenBar95}. These classical mean-field limits rely on a strong form of redundancy: the existence of large ensembles of neuronal duplicates receiving (almost) the same recurrent and external input. 
    %To our knowledge, this form of redundancy has not been found in cortex.
    \item \textit{Temporal averaging over single-neuron spikes}: In~\eqref{eq:SNN}, we can replace the transfer function $\phi$ and the weights $\{J_{ij}\}_{i\neq j}$ by $b\,\phi$ and $\{J_{ij}/b\}_{i\neq j}$, respectively (for $b>0$), and take the limit $b\to\infty$ \cite{Kur71}. This limit entails arbitrarily high firing rates in the SNN, which is biologically unrealistic since two spikes have to be separated by at least $1$ to $2$ milliseconds (the absolute refractory period) \cite{RieWar97, KanSCh00}. Alternatively, but to a similar effect, we can take the limit $\tau \to \infty$ in both~\eqref{eq:SNN} and~\eqref{eq:RNN} while re-scaling the weights $\{J_{ij}\}_{i\neq j}$ and the external inputs $I^\mathrm{ext}_i(t)$ by $1/\tau$. This last limit entails arbitrarily slow network dynamics, which is incompatible{\color{black}, for example,} with human visual processing speed (less than $150$ milliseconds) \cite{ThoFiz96}. 
\end{enumerate}

\begin{figure*}
    \includegraphics[width=\textwidth]{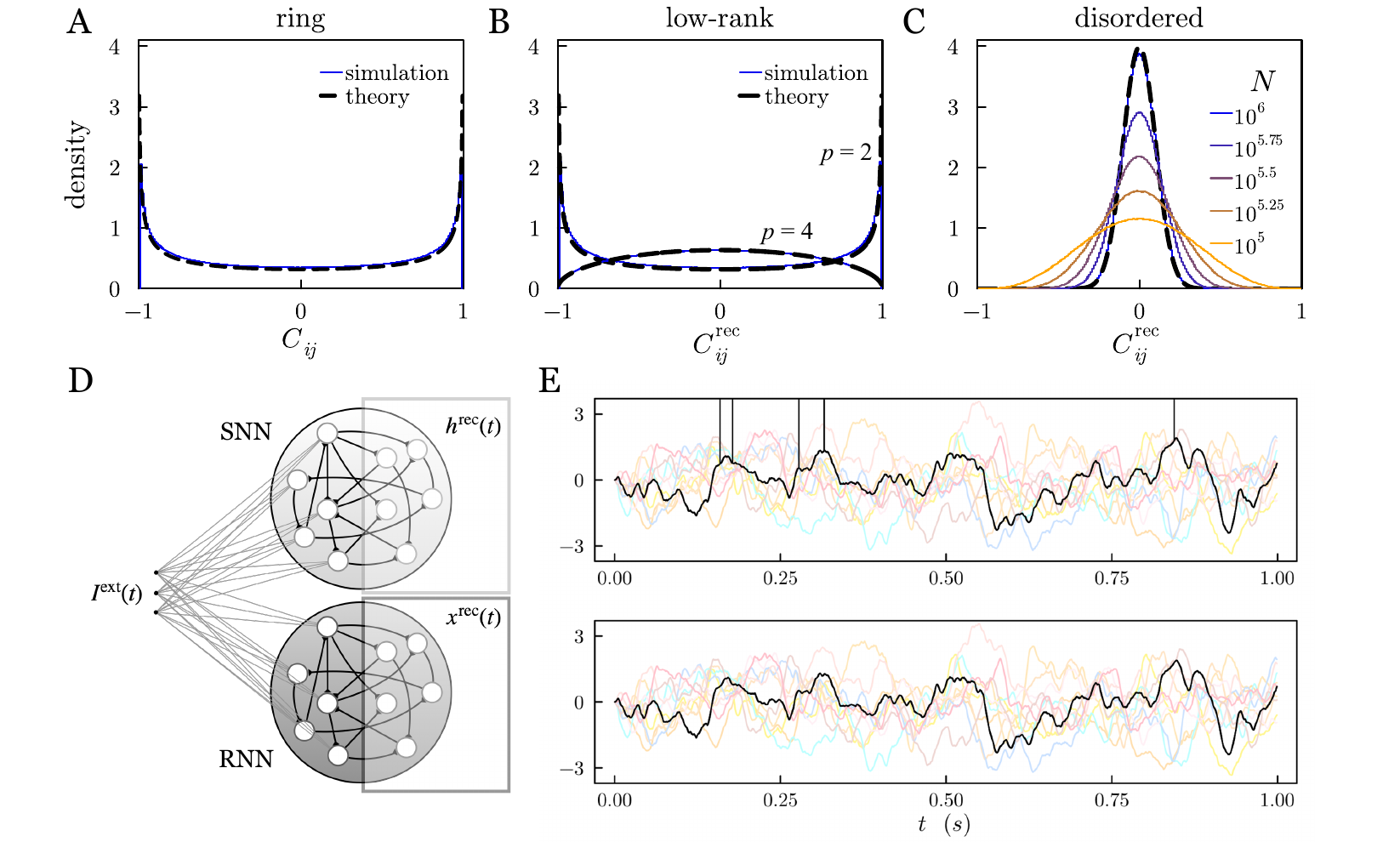}
    \caption{\label{fig:setup} (\textbf{A}) Limit correlation density for the ring model. The histogram of a simulation with $N=10^6$ units (solid blue line) is compared to the theoretical limit density~\eqref{eq:theory_ring} (dashed black line).
    (\textbf{B}) Limit correlation density (of units receiving recurrent input only) for low-rank networks $p=2$ and $p=4$. Simulations of RNNs with $N=10^6$ units (solid blue line) are compared to the corresponding theoretical limit ``Gegenbauer'' density~\eqref{eq:gegenbauer} (dashed black line).
    (\textbf{C}) Concentration around $0$ of the correlation distribution in disordered networks with $p=\alpha N$ and fixed load $\alpha=10^{-4}$. The correlation distribution (of units receiving recurrent input only) of simulations of networks of increasing size $N$ converge to a centred normal density with variance $1/p=1/(\alpha N)$ (the dashed black line indicates the normal density for $N=10^6$).
    (\textbf{D}) Setup for comparing SNNs and RNNs.
    %SNN and RNN comparison. Only half of the neurons in the SNN or the RNN receive external input $I^{\mathrm{ext}}(t)$.
    %The potentials of the neurons receiving no external input but only recurrent input, $h^{\mathrm{rec}}(t)$ and $x^{\mathrm{rec}}(t)$, are linear readouts of the recurrent drive. 
    (\textbf{E}) Trajectories of single-neuron potentials in the SNN ($h^{\mathrm{rec}}_i(t)$, upper panel) and in the RNN ($x^{\mathrm{rec}}_i(t)$, lower panel) during a one-second simulation of the setup shown in (D). The networks have $N=10^6$ neurons and load $\alpha = 10^{-4}$, as in (C). The same randomly chosen $11$ neurons are recorded in the SNN and in the RNN and each colour corresponds to a different neuron (the colours in the upper and lower panels correspond). For one neuron (the black trace), the spike times of the neuron in the SNN are indicated by vertical bars. 
    %The differences between the potentials of the SNN and the RNN are almost imperceptible. 
    (\textbf{B}-\textbf{E}) {\color{black}Model parameters: $\tau=10$~ms and $\sigma=0.5$ (see~\eqref{eq:input})}.}
\end{figure*}

In this letter, we address the following question: can large SNNs, as defined in~\eqref{eq:SNN}, converge to {\color{black}\textit{equally large}} RNNs in the absence of neuronal duplicates and without temporal averaging?

Temporal averaging as in (ii) is made impossible if we impose, as we do, that $\max \phi \leq 1/\tau$. Under this constraint, leaky integration by the potential~\eqref{eq:SNN} is too fast to average out the Poisson noise of individual input spike trains and neither of the two scalings mentioned under (ii) can be applied.

To quantify the amount of duplication in an RNN, we look at the distribution of correlations \cite{RenDel10} between pairs of distinct neurons $i\neq j$
\begin{equation}\label{eq:correl_def}
    C_{ij} := \lim_{T \to \infty}\frac{1}{T}\int_0^T \frac{(x_i(t) - \overline{x_i})(x_j(t) - \overline{x_j})}{\sigma_{x_i}\sigma_{x_j}}dt,
\end{equation}
where $\overline{x_i}$ and $\sigma_{x_i}^2$ are, respectively, the time average and fluctuation of $x_i(t)$ ($\overline{x_i} := \lim_{T \to \infty}\frac{1}{T}\int_0^T x_i(t)dt$ and $\sigma_{x_i}^2 := \lim_{T \to \infty}\frac{1}{T}\int_0^T (x_i(t) - \overline{x_i})^2dt$.
%\begin{equation}
%\begin{aligned}\label{eq:stationary}
%    \overline{x_i} &:= \lim_{T \to \infty}\frac{1}{T}\int_0^T x_i(t)dt, \\
%    \sigma_{x_i}^2 &:= \lim_{T \to \infty}\frac{1}{T}\int_0^T (x_i(t) - \overline{x_i})^2dt.
%\end{aligned}
%\end{equation}

To illustrate how duplicates accumulate in classical mean-field models, let us consider a toy example with a ring structure. Let the $N$ units of an RNN be uniformly and independently positioned on a ring, $\theta_i$ denoting the angular position of unit $i$. If the correlation between unit $i$ and $j$ is given by $C_{ij} := \cos(\theta_i - \theta_j)$, then, as $N\to\infty$, the limit distribution of the pairwise correlation, which we call the limit correlation density, is 
\begin{equation}\label{eq:theory_ring}
    \rho(z) = \mathbbm{1}_{[-1,1]}(z)\frac{1}{\pi}\frac{1}{\sqrt{1-z^2}}
\end{equation}
(Fig~\ref{fig:setup}A; proof in {\color{black}SM~Sec.~I }\footnote{See Supplemental Material at [URL will be inserted by publisher]}). This limit density has a strictly positive mass around $1$, reflecting the fact that each unit has a number of duplicates that grows linearly with $N$.

We say that large networks are \textit{duplicate-free} if, for any threshold $\varepsilon>0$, the probability that there exists a pair of distinct neurons $i\neq j$ such that their absolute correlation is greater or equal to $\varepsilon$ tends to $0$ as $N\to\infty$, {\color{black}i.e. for all $\varepsilon>0$,
\begin{equation}\label{eq:def_duplicate_free}
    \mathbb{P}\left(\exists\, (i,j) \text{ with }i\neq j \text{ s.t. } |C_{ij}|\geq \varepsilon\right) \xrightarrow[N\to\infty]{} 0.
\end{equation}}
In the following, we propose a disordered network model where large networks are duplicate-free. We construct the connectivity matrix $\J = \{J_{ij}\}_{i,j}$ as a sum of random rank-one matrices (minus self-interaction terms), a construction similar to that of Hopfield networks {\color{black}\cite{Ama72, Hop82,AmiGut85a,Her89,PerBru18}}. For any number of units $N$ and any number of patterns $p$, let $\bxi$ be a random $N \times p$-matrix with \textit{i.i.d.} {\color{black}standard normal} entries $\{\xi_{i\mu}\}_{i,\mu}$. {\color{black}The }connectivity matrix {\color{black}is }given by
\begin{equation}\label{eq:struct}
    J_{ij} := \frac{1}{cN}\sum_{\mu=1}^p \xi_{i\mu}\left(\phi(\xi_{j\mu}) - a\right) \quad \text{for all }i\neq j,
\end{equation}
and $J_{ii} := 0$ for all $i$, where $a:=\int_{-\infty}^\infty \mathcal{D}z\; \phi(z)$ and $c:=\int_{-\infty}^\infty \mathcal{D}z\; (\phi(z) - a)^2$ are fixed constants ($\mathcal{D}z$ denotes the standard Gaussian measure). {\color{black}The nonlinearity $\phi(x) := \frac{1}{2\tau}\left(\tanh(x-b)+1\right)$, with $b=2$, is chosen such that the autonomous dynamics of the RNN is in the so-called ``paramagnetic'' \cite{KuhBos91} or ``background'' \cite{PerBru18} state, where all units have a low firing rate.}
%guarantee the normalization
%\begin{equation*}
%    \frac{1}{cN}\sum_{i=1}^N\left(\phi(\xi_{i1}) - a\right)\phi(\xi_{i1}) \to 1, \quad \text{as }N \to\infty.
%\end{equation*}

A well-known feature of this type of connectivity is that, exchanging the order of summation, the dynamics of the SNN~\eqref{eq:SNN} can be re-written in terms of $p$ {\color{black}latent factors} $\{m_\mu(t)\}_{\mu}$ {\color{black}\cite{Ami89,KuhBos91,GerKis14}}: for all $i=1,\dots,N$ and for all $\mu=1,\dots,p$,
\begin{align}
    \tau \frac{\rd}{\rd t}h_i(t) &= - h_i(t) + \sum_{\nu=1}^p \xi_{i\nu}m_\nu(t) - \gamma_i S_i(t)+ I^{\mathrm{ext}}_i(t), \nonumber\\
    m_\mu(t)&=\frac{1}{cN}\sum_{j=1}^N\left(\phi(\xi_{j\mu}) - a\right)S_j(t). \label{eq:overlap}
\end{align}
%\begin{align*}
%    \tau \frac{\rd}{\rd t}x_i(t) &= - x_i(t) + \sum_{\nu=1}^p \xi_{i\nu}m_\nu(t) - \gamma_i \phi(x_i(t))+ I^{\mathrm{ext}}_i(t), \nonumber\\
%    m_\mu(t)&=\frac{1}{cN}\sum_{j=1}^N\left(\phi(\xi_{j\mu}) - a\right)\phi(x_j(t)),
%\end{align*}
where the $\gamma_i:= \frac{1}{cN}\sum_{\nu=1}^p\xi_{i\nu}\left(\phi(\xi_{i\nu})-a\right)$ are virtual self-interaction weights. An analogous reformulation holds for the RNN~\eqref{eq:RNN}. The reformulation clearly shows that if $p \ll N$, the $\gamma_i$ are small and therefore the recurrent drive $\{\sum_{j=1}^N J_{ij}S_j(t)\}_i$ is approximately restricted to the $p$-dimensional subspace spanned by the $p$ columns of the random matrix $\bxi$. To force the recurrent drive to visit all $p$ {\color{black}dimensions over time}, we inject the following $p$-dimensional external input to half of the neurons:
\begin{equation}
\begin{aligned}\label{eq:input}
    I^{\mathrm{ext}}_i(t) &= \frac{\sigma}{\sqrt{p}}\sum_{\mu=1}^p \xi_{i\mu}\eta_\mu(t) &&\text{if }i\leq N/2,\\
    %I^{\mathrm{ext}}_i(t) &= 0 &&\text{if }i> N/2,
\end{aligned}
\end{equation}
where the $\eta_1(t), \dots, \eta_p(t)$ are {\color{black}the formal derivatives of independent standard Brownian motions $B_1(t), \dots, B_p(t)$, i.e. $\eta_\mu(t) = \rd B_\mu(t)/\rd t$, }and $\sigma>0$ is the standard deviation of the input. For clarity, we use the superscript `in' to emphasize that, for all $i\leq N/2$, the neurons $h^{\mathrm{in}}_i(t)$ and the units $x^{\mathrm{in}}_i(t)$ receive external as well as recurrent inputs, and we use the superscript `rec' to emphasize that, for all $i>N/2$, the neurons $h^{\mathrm{rec}}_i(t)$ and the units $x^{\mathrm{rec}}_i(t)$ receive recurrent input only. {\color{black}Henceforth, we write $C^{\mathrm{rec}}_{ij}$ for the pairwise correlation~\eqref{eq:correl_def} between the `rec' units $x^{\mathrm{rec}}_i$ and $x^{\mathrm{rec}}_j$.
Assuming that, for any $N$ and $p$, the time-dependent input \eqref{eq:input} leads to a stationary ergodic state where the recurrent drive is restricted to and visits all $p$ dimensions given by the columns of $\bxi$ homogeneously, we make the following approximation
\begin{equation}\label{eq:bound_correl}
    C^{\mathrm{rec}}_{ij} \approx \frac{\sum_{\mu=1}^p \xi_{i\mu}\xi_{j\mu}}{\sqrt{\sum_{\mu=1}^p \xi_{i\mu}^2}{\sqrt{\sum_{\mu=1}^p \xi_{j\mu}^2}}}.
\end{equation}
Although the approximation~\eqref{eq:bound_correl} is heuristic, it proves to by highly accurate in the context of the models considered here, as demonstrated by simulations (Fig~\ref{fig:setup}B,C).
}
%we find, for any $N$ and $p$, that the correlation between two units satisfies the bound
%\begin{equation}\label{eq:bound_correl}
%    C_{ij} \leq \frac{\sum_{\mu=1}^p \xi_{i\mu}\xi_{j\mu}}{\sqrt{\sum_{\mu=1}^p \xi_{i\mu}^2}{\sqrt{\sum_{\mu=1}^p \xi_{j\mu}^2}}},
%\end{equation}
%because of the radial symmetry of the joint stationary distribution of the variables $\{m_\mu\}_{\mu=1}^p$.
%Moreover, for any two `rec' units (or any two `in' units), we have the approximation
%\begin{equation}\label{eq:correl}
%    \rho(x^{\mathrm{rec}}_i,x^{\mathrm{rec}}_j) \approx \frac{\sum_{\mu=1}^p \xi_{i\mu}\xi_{j\mu}}{\sqrt{\sum_{\mu=1}^p \xi_{i\mu}\xi_{i\mu}}{\sqrt{\sum_{\mu=1}^p \xi_{j\mu}\xi_{j\mu}}}}.
%\end{equation}
%The bound~\eqref{eq:bound_correl} is tight, if the virtual self-interaction weights $\gamma_i=0$, that is, if $J_{ii}=\frac{1}{cN}\sum_{\mu=1}^p\xi_{i\mu}\left(\phi(\xi_{i\mu})-a\right)$. 
%The bound~\eqref{eq:bound_correl} also holds for the pairwise correlations in the SNN~\eqref{eq:SNN} since the addition of spike noise can only reduce correlations. 
%Henceforth, we write $C^{\mathrm{rec}}_{ij}$ {\color{black}for} the pairwise correlation~\eqref{eq:correl_def} between the `rec' units $x_i$ and $x_j$.

If the number of patterns $p$ is kept constant as $N\to\infty$, the limit RNN is a low-rank mean-field model (without disorder) \cite{MasOst18,BeiDub21}. 
%In such low-rank models, the bound~\eqref{eq:bound_correl} is asymptotically tight for the $C^{\mathrm{rec}}_{ij}$, since the virtual self-interaction weights $\gamma_i\to 0$ as $N\to\infty$. 
{\color{black}In this case}, duplicates accumulate as $N\to\infty$ because the distribution of correlations $\{C^{\mathrm{rec}}_{ij}\}_{i<j}$ {\color{black}(under the approximation~\eqref{eq:bound_correl})} converges to a limit density with a strictly positive mass around $1$, indicating that the number of duplicates per unit grows linearly with $N$. More specifically, 
%by the Funk-Hecke formula, 
for any fixed $p>1$, the limit density is given by the so-called ``Gegenbauer distribution'' with parameter $p$, 
\begin{equation}\label{eq:gegenbauer}
    \rho(z) = \mathbbm{1}_{[-1,1]}(z)\frac{\Gamma(p/2)}{\sqrt{\pi}\,\Gamma((p-1)/2)}(1-z^2)^{\frac{p-3}{2}},
\end{equation}
which is the orthogonal projection of the uniform distribution on the unit sphere $\mathbb{S}^{p-1}$ onto its diameter
(Fig~\ref{fig:setup}B; proof in {\color{black}SM~Sec.~I} \footnotemark[1]). Intuitively, the explanation for the accumulation of duplicates is the same as in the ring model presented above, except that here the fixed space is not a ring but $\mathbb{R}^p$: unit $i$ has {\color{black}random normal coordinates} $\bxi_i = (\xi_{i1}, \dots, \xi_{ip})$ {\color{black}\cite{HemGre86,HemGre88}} and units with similar coordinates receive similar recurrent and external inputs. Therefore, if $p$ is kept fixed as $N\to\infty$, we fall into the case of spatial averaging over neuronal duplicates (i) leading to neural field equations (see {\color{black}SM~Sec.~II} \footnotemark[1]).
%\begin{equation*}
%    \lim_{N\to\infty}h_i(t) = \lim_{N\to\infty}x_i(t) = u(t,\bxi_i),
%\end{equation*}
%where $u(t,\by)$ is the solution to the integro-differential equation
%\begin{align*}
%    \tau \frac{\partial}{\partial t}u(t,\by) &= -u(t,\by) + \int_{\R^p}\mathcal{D}\bz \;w(\by,\bz)\phi(u(t,\bz)), \\
%    w(\by,\bz) &= \frac{1}{c}\sum_{\mu=1}^p y_\mu \left(\phi(z_\mu) - a\right),%+ \frac{1}{c}\sum_{\mu=1}^p y_\mu\int_{\R^p}\mathcal{D}\bz\,\left(\phi(\bz) - a\right)\phi(u(t,\bz)),
%\end{align*}
%where $\mathcal{D}\bz$ is the $p$-dimensional standard Gaussian measure.

To prevent duplicates from accumulating as $N\to\infty$, we make the number of patterns $p$ grow linearly with $N$, taking $p=\alpha N$ for some fixed load $\alpha>0$. With this choice of scaling, weights $\{J_{ij}\}_{i,j}$ scale as $\mathcal{O}(1/\sqrt{N})$ (as in random RNNs \cite{SomCri88, KadSom15, VanAlb21, ClaAbb23}). First~(I), we will show that for any fixed $\alpha>0$, large networks are \textit{duplicate-free}. Second~(II), we will show that large SNNs converge to large RNNs, with convergence rate $\sqrt{\alpha}$, as $\alpha \to 0$ (this means that for arbitrarily small $\alpha>0$, large SNNs behave almost exactly like large RNNs).
%achieving the numerical demonstration -- supported by theory -- that large, duplicate-free SNNs can converge to equally large RNNs. 
To compare the SNN~\eqref{eq:SNN} with the RNN~\eqref{eq:RNN}, we will inject the same time-dependent external input~\eqref{eq:input} in both networks (Fig~\ref{fig:setup}D) and compare the trajectories $h^\mathrm{rec}_i(t)$ of the SNN with the trajectories $x^\mathrm{rec}_i(t)$ of the RNN (Fig~\ref{fig:setup}E).

%We first verify numerically that the SNNs (and the RNNs) converge to the dynamic mean-field limit of a disordered system \cite{SomCri88} (not to be confused with classical mean-field limits where there is no disorder \cite{Ger95,Ger00}). Since the trajectories $h^\mathrm{rec}_i(t)$ of the SNN are almost identical to the trajectories $x^\mathrm{rec}_i(t)$ of the RNN when $\alpha$ is small ($\alpha=10^{-4}$ in Fig.~\ref{fig:setup}C,E and Fig~\ref{fig:rec_drive}A-E), we only show numerical results for the SNN; those for the RNN are almost identical. Defining the time average $\overline{h^\mathrm{rec}_i}$ and the fluctuation $\sigma_{h^\mathrm{rec}_i}$ as in Eq.~\eqref{eq:stationary}, we see that the distributions of $\overline{h^{\mathrm{rec}}_i}$ concentrate around $0$ (Fig~\ref{fig:rec_drive}A) and the distributions of $\sigma_{h^{\mathrm{rec}}_i}$ concentrate around a value slightly larger than $1$ (Fig~\ref{fig:rec_drive}B) as $N\to\infty$. Moreover, when $N$ is large, the trajectories $h^\mathrm{rec}_i(t)$ (and $x^\mathrm{rec}_i(t)$) look like independent realizations of the same stochastic process (Fig.~\ref{fig:setup}E). While the comprehensive study of this putative dynamic mean-field theory is beyond the scope of this letter, a feedforward network approximation gives a good intuition for the limit stationary dynamics \footnote{See Supplementary Materials ``Feedforward network approximation''}.

\begin{figure}
    \includegraphics[width=\columnwidth]{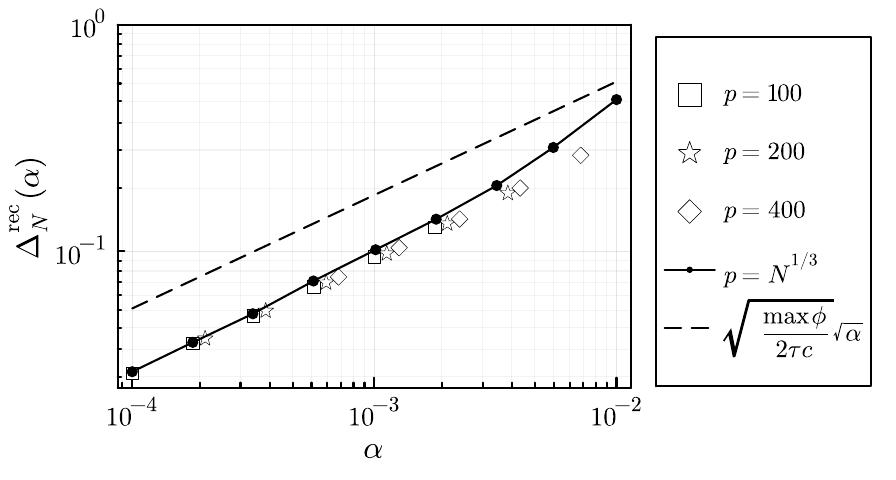}
    \caption{\label{fig:2} Large $N$ simulations of the distance $\Delta^\mathrm{rec}_N(\alpha)$ as a function of $\alpha$ for $p=100$ (rectangles), $p=200$ (stars), and $p=400$ (diamonds). 
    Simulations of $\Delta^\mathrm{rec}_N(\alpha)$ for the duplicate-free sequence of networks $p=N^{1/3}$ (full circles).
    Theoretical $\mathcal{O}(\sqrt{\alpha})$ bound predicated by the feedforward model simplification (dashed line).
    %(\textbf{A}) Distributions of single-neuron distances $\Delta^\mathrm{rec}_i$ for increasing network size $N$ and fixed load $\alpha=10^{-4}$. Distances concentrate around a limit distance $\Delta(\alpha)\approx 0.03$ as $N\to\infty$ (dashed cyan line).
    %(\textbf{B}) Numerical estimate of the limit distance $\Delta(\alpha)$ for various fixed-load $\alpha$ and fitted power law with exponent $1/2$ (dashed line). The value for $\alpha=10^{-4}$ is indicated by a cyan circle.
    Same parameters as in Fig.~\ref{fig:setup}B-E.}
\end{figure}

\textit{(I) Large networks are duplicate-free}.--- {\color{black}Under the approximation~\eqref{eq:bound_correl}, by the central limit theorem, we have that,} for any distinct pair of units $i\neq j$, the scaled correlation $\sqrt{p}\,C^{\mathrm{rec}}_{ij}$ converges in law to a standard normal random variable as $p\to\infty$. %, i.e.
%\begin{equation*}
%    \sqrt{p}\,\rho(x^{\mathrm{rec}}_i,x^{\mathrm{rec}}_j)\xrightarrow[p\to\infty]{\mathcal{L}} X, \quad X\sim\mathcal{N}(0,1).
%\end{equation*}
Then, if $p\to\infty$ as $N\to\infty$, we can prove that the distribution of correlations $\{C^{\mathrm{rec}}_{ij}\}_{i<j}$ converges to a centered normal distribution with variance $1/p$ as $N\to\infty$ (Fig.~\ref{fig:setup}C; proof in {\color{black}SM~Sec.~I} \footnotemark[1]). {\color{black}Moreover, using the fact that for all $N,p>1$ and $i\neq j$, $C^{\mathrm{rec}}_{ij}$ follows a} Gegenbauer distribution with parameter $p$~\eqref{eq:gegenbauer}, we can derive a bound for the probability of a duplicate:
\begin{multline}\label{eq:bound_probab}
    \mathbb{P}\left(\exists\, (i,j) \text{ with }i\neq j \text{ s.t. } |{\color{black}C^{\mathrm{rec}}_{ij}}| \geq \varepsilon\right) \\
    \leq \frac{N(N-1)}{2}\frac{(1 -\varepsilon^2)^{\frac{p-1}{2}}}{\sqrt{\pi}},
\end{multline}
for all $0<\varepsilon<1$ (see {\color{black}SM~Sec.~III} \footnotemark[1]). Since $p=\alpha N$, the bound tends to $0$ as $N\to\infty$, which confirms that large networks are duplicate-free {\color{black}(\eqref{eq:def_duplicate_free} can be easily deduced from \eqref{eq:bound_probab})}.

\textit{(II) Large SNNs converge to large RNNs at rate $\sqrt{\alpha}$, as $\alpha \to 0$}.--- For any fixed load $\alpha$, we define the average distance between the SNN and the RNN as 
\begin{equation*}
    \Delta^\mathrm{rec}_N(\alpha) := \frac{2}{N}\sum_{i=N/2+1}^{N}\lim_{T\to\infty}\frac{1}{T}\int_0^T\left|h^{\mathrm{rec}}_i(t) - x^{\mathrm{rec}}_i(t)\right|\rd t 
\end{equation*}
and the large $N$ limit distance as $\Delta^\mathrm{rec}(\alpha):=\lim_{N\to\infty}\Delta^\mathrm{rec}_N(\alpha)$.
%Let us first observe that, for a fixed $\alpha>0$ and as $N\to\infty$, the distributions of the single-neuron distances
%\begin{equation*}
%    \Delta^\mathrm{rec}_i := \lim_{T\to\infty}\frac{1}{T}\int_0^T\left|h^{\mathrm{rec}}_i(t) - x^{\mathrm{rec}}_i(t)\right|\rd t. 
%\end{equation*}
%concentrate around a finite value (around $0.03$ for $\alpha=10^{-4}$, Fig.~\ref{fig:2}A). 
%We define, for any $\alpha>0$, the $N\to\infty$ limit distance between the SNN and the RNN as the average of the single-neuron distances
%\begin{equation}\label{eq:Delta_def}
%    \Delta^\mathrm{rec}(\alpha) := \lim_{N\to\infty}\frac{2}{N}\sum_{i=1}^{N/2}\Delta^\mathrm{rec}_i.
%\end{equation}
Numerical estimates of the limit distance $\Delta^\mathrm{rec}(\alpha)$ as a function of $\alpha$ indicates that
\begin{equation}\label{eq:Delta}
    \Delta^\mathrm{rec}(\alpha)= \mathcal{O}(\sqrt{\alpha}), \quad\text{as }\alpha\to 0
\end{equation}
(Fig.~\ref{fig:2}). This scaling implies that large SNNs can converge to equally large RNNs despite the fact that (i) there is no duplicate averaging (Fig.~\ref{fig:setup}C and~\eqref{eq:bound_probab}) and (ii) no temporal averaging. For example, if we take the sublinear scaling $p=N^{1/3}$, which implies $\alpha=N^{-2/3}$, the distance $\Delta^\mathrm{rec}_N(\alpha)\to 0$ vanishes {\color{black}as $N\to\infty$} (Fig.~\ref{fig:2}, full circles) and {\color{black}large }networks are duplicate-free~\eqref{eq:bound_probab}.

%More concretely, this result suggest that for any increasing sequence $p(N)$ such that both the bound on the probability of a duplicate~\eqref{eq:bound_probab} and $\alpha=p/N$ converge to $0$ as $N\to\infty$, the SNNs will converge to the corresponding RNNs even though the networks are duplicate-free in the limit. 

%The concentration of measure phenomenon \cite{Tal96} does implicitly rely on averaging of types (i) and (ii): our disordered network model gives a concrete illustration of this fact.
%Our network model gives a concrete example where, even in the absence of averaging of types (i) and (ii), the concentration of measure phenonemon (`a random variable that depends on many independent variables but not too much on any of them is essentially constant' \cite{Tal96}) suffices for SNNs to converge to equally large RNNs. 
In the absence of averaging of types (i) and (ii), the concentration of measure phenomenon \cite{Tal96, Led01} can explain the convergence of SNNs to RNNs. In our case, the concentration of measure is controlled by the $\ell^2$ norm $\|\J_i\|_2 = \sqrt{\sum_{j=1}^N J_{ij}^2}$ of each neuron $i$'s incoming weights. When $p=\alpha N$, for a typical neuron $i$, we find the convergence in probability 
\begin{equation}\label{eq:weight_convergence}
    \|\J_{i}\|_2 \xrightarrow[N\to\infty]{\mathbb{P}}\sqrt{\frac{\alpha}{c}},
\end{equation}
(proof in {\color{black}SM~Sec.~IV} \footnotemark[1]) which means that the $\ell^2$ norms of each neuron's incoming weights concentrate around $\sqrt{\alpha/c}$ (where $c$ is defined after~\eqref{eq:struct}). The numerical scaling \eqref{eq:Delta} of the limit distance between large SNNs and large RNNs, as $\alpha \to 0$, corresponds to the theory-based scaling \eqref{eq:weight_convergence} of the limit $\ell^2$ norm of a typical neuron's incoming weights. 

\begin{figure}
    \includegraphics[width=\columnwidth]{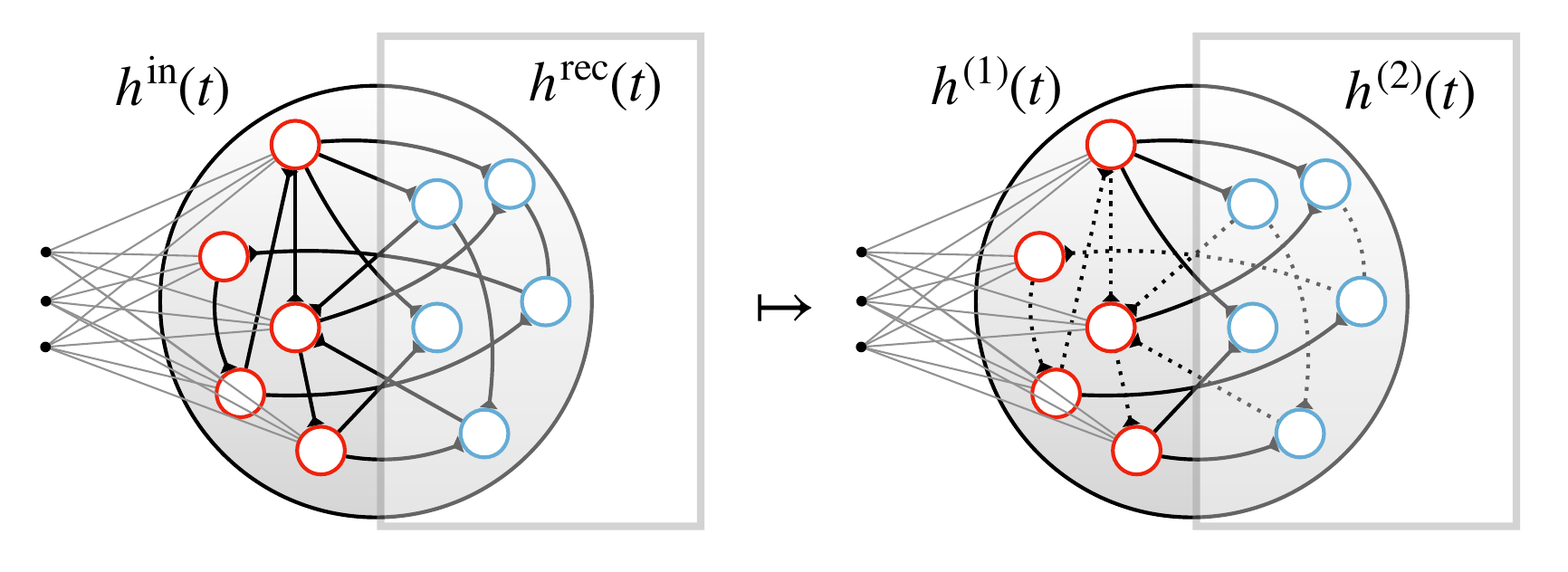}
    \caption{\label{fig:main_text_ff}The feedforward model as a simplification of the input-driven model. Dotted lines indicate removed connections}
\end{figure}

Although we do not have an exact theory linking the limit $\ell^2$ norm~\eqref{eq:weight_convergence} with the limit distance $\Delta^{\mathrm{rec}}(\alpha)$~\eqref{eq:Delta}, a simplified feedforward model, which is analytically tractable, offers {\color{black}an} intuition for the concentration of measure phenomenon at play. This simplified model is obtained by keeping network connections from `in' to `rec' neurons and removing all the other connections (Fig.~\ref{fig:main_text_ff}). 
After this pruning procedure, we are left with two-layer feedforward networks where the `in' neurons make up the first layer, $(h^{\mathrm{in}}_i(t), x^{\mathrm{in}}_i(t)) \mapsto (h^{(1)}_i(t), x^{(1)}_i(t))$, and the `rec' neurons make up the second layer, $(h^{\mathrm{rec}}_i(t), x^{\mathrm{rec}}_i(t)) \mapsto (h^{(2)}_i(t), x^{(2)}_i(t))$. 
%We study the convergence of $h^{(2)}_i(t)$ to $x^{(2)}_i(t)$ as a proxy for the original question of the convergence of the SNNs' $h^{\mathrm{rec}}_i(t)$ to the RNNs' $x^{\mathrm{rec}}_i(t)$. 
In general, for any $N>1$ and for \textit{any} connectivity matrix $\J$, we find that, for any neuron $i$ in the second layer, the single-neuron distance satisfies the bound
\begin{equation*}
    \lim_{T\to\infty}\frac{1}{T}\int_0^T\left|h^{(2)}_i(t) - x^{(2)}_i(t)\right|\rd t \leq \sqrt{\frac{\max\phi}{2\tau}} \|\J_i\|_2,
\end{equation*}
(proof in {\color{black}SM Sec.~V} \footnotemark[1]) which is indeed controlled by the $\ell^2$ norm $\|\J_i\|_2$. Then, if $p=\alpha N$, using the convergence of the $\ell^2$ norm~\eqref{eq:weight_convergence}, we get the expected scaling for the limit distance, as $N\to\infty$, between a spiking neuron $h^{(2)}_{i}$ and a rate unit $x^{(2)}_{i}$: 
\begin{equation*}
\lim_{T\to\infty}\frac{1}{T}\int_0^T\left|h^{(2)}_{i}(t) - x^{(2)}_{i}(t)\right|\rd t \leq \sqrt{\frac{\max\phi}{2\tau c}}\sqrt{\alpha}.
\end{equation*}
%\begin{equation*}
%    \lim_{T\to\infty}\frac{1}{T}\int_0^T\left|h^{(2)}_{i^*}(t) - x^{(2)}_{i^*}(t)\right|dt = \mathcal{O}(\sqrt{\alpha}).
%\end{equation*}
%(cf. Fig.~\ref{fig:2}). Therefore, the feedforward model provides an intuition for how the vanishing $\ell^2$ norms of the incoming weights (as $\alpha \to 0$) cause concentration of measure in duplicate-free, large SNNs. An exact theory of the original recurrent networks would require a full-fledged dynamical mean-field theory \cite{HelDah20} for our input-driven disordered network model, which remains an open problem.

Concentration of measure \cite{Led01} has been shown to be instrumental for the theory of spin glasses \cite{Tal10} and the theory of artificial neural networks \cite{MeiMon22}. In contrast, this probabilistic notion has barely permeated the theory of {\color{black}biological neural networks}. The standard perspective has been that, to produce spike-noise-robust population dynamics, large networks have to perform averages over the spike activity of many neuronal duplicates \cite{ShaNew98,GerKis02,GerKis14,Bre15}, i.e. mechanism (i). 
%This reasoning stems from the law of large numbers, which is at the core of classical mean-field theories \cite{WilCow72,WilCow73,Ger95}. 
Recently, \citet{DepSus23} proposed the perspective that ``weighted averages'' over heterogeneous neurons could lead to population-level latent factors with noise-robust dynamics, which then produce the illusion of rate-based dynamics. {\color{black}We provide a theoretical foundation for this idea by showing that the concentration of measure phenomenon explains how the dynamics of large, duplicate-free SNNs can converge to the dynamics of equally large RNNs. {\color{black}This convergence does not require symmetric weights and holds even when diffusive membrane noise is added to the potential dynamics (SM~Secs.~VI,~VII~\footnotemark[1]).}
%Note that, while we have focused our analysis on networks of linear-nonlinear-Poisson neurons, analogous results {\color{black}still hold, for example, when diffusive membrane noise is added to the potential dynamics} (see SM~Sec.~VI \footnotemark[1]). 
In the absence of neuronal duplicates, rate-based population dynamics in networks of spiking neurons can only be understood as an emergent behaviour: neurons interact {\color{black}through} spikes, but they behave \textit{as if} they were interacting {\color{black}through} their unique, time-varying firing rates. This notion of emergent rate-based dynamics sheds new light on the long-standing spike vs. rate debate in computational neuroscience \cite{ShaNew98,GerKis02,Bre15}.}

\begin{acknowledgments}
This research is supported by the Swiss National Science Foundation (no 200020\_207426). V.S. is also supported by a Royal Society Newton International Fellowship (NIF$\backslash$R1$\backslash$231927). The authors thank Juhan Aru for a useful discussion on the proof of Theorem~2 in the Supplemental Materials, as well as Louis Pezon and Nicole Vadot for their valuable comments on the manuscript.
\end{acknowledgments}

\bibliography{ms}
\end{document}

%\preprint{}

%\title{Convergence of duplicate-free spiking neural networks to rate networks}
%\title{Convergence of duplicate-free spiking neural networks to rate networks}
%\title{Convergence of duplicate-free spiking neural networks to rate-based dynamics}
\title{Supplementary Material on \\
Emergent rate-based dynamics in duplicate-free populations of spiking neurons}
%\title{From spike-based interaction to rate-based population dynamics: concentration of measure in duplicate-free networks}
\author{Valentin Schmutz}
\email[Corresponding author: ]{v.schmutz@ucl.ac.uk} 
\affiliation{School of Life Sciences and School of Computer and Communication Sciences, École Polytechnique Fédérale de Lausanne, 1015 Lausanne, Switzerland}
\affiliation{UCL Queen Square Institute of Neurology, University College London, WC1E 6BT London, United Kingdom}
\author{Johanni Brea}
\author{Wulfram Gerstner}
%\author{Valentin Schmutz\footnote{Corresponding author: v.schmutz@ucl.ac.uk}, Johanni Brea, Wulfram Gerstner\footnote{Corresponding author: wulfram.gerstner@epfl.ch}}
\affiliation{School of Life Sciences and School of Computer and Communication Sciences, École Polytechnique Fédérale de Lausanne, 1015 Lausanne, Switzerland}

%%%%%%%%%% Prefix a "S" to all equations, figures, tables and reset the counter %%%%%%%%%%
\setcounter{equation}{0}
\setcounter{figure}{0}
%setcounter{table}{0}
%\setcounter{page}{1}
%\makeatletter
\renewcommand{\theequation}{S\arabic{equation}}
\renewcommand{\thefigure}{S\arabic{figure}}

\maketitle
\tableofcontents

\ensureoddstart
\section{Limit correlation density}
We prove two theorems characterizing the limit distribution of pairwise correlations between distinct {\color{black}units} $\{C_{ij}\}_{i\neq j}$, as $N\to\infty$, assuming that the correlation $C_{ij}$ between a pair of distinct {\color{black}units} $i\neq j$ is given by
\begin{equation}\label{eq:corr_def_appendix}
    C_{ij} = \frac{1}{p}\sum_{\mu=1}^p \tilde{\xi}_{i\mu}\tilde{\xi}_{j\mu},
\end{equation}
where $\tilde{\xi}_{i\mu} := \xi_{i\mu}/\sqrt{\frac{1}{p}\sum_{\mu=1}^p \xi_{i\mu}^2}$ and the $\{\xi_{i\mu}\}_{1\leq i \leq N, 1\leq \mu \leq p}$ are $i.i.d.$, zero-mean, unit-variance normal random variables {\color{black}(see Approximation~(9) in the main text)}. More explicitly, we prove the convergence of the random empirical measure
\begin{equation}\label{eq:empirical}
    \rho_N(\rd \bz):=\frac{2}{N(N-1)} \sum_{1\leq i < j \leq N}\delta_{\frac{1}{p}\sum_{\mu=1}^p \tilde{\xi}_{i\mu}\tilde{\xi}_{j\mu}}(\rd z),
\end{equation}
to a deterministic probability measure as $N\to\infty$. In \eqref{eq:empirical}, $\delta_{\frac{1}{p}\sum_{\mu=1}^p \tilde{\xi}_{i\mu}\tilde{\xi}_{j\mu}}(\rd z)$ denotes the Dirac measure centered on the point $\frac{1}{p}\sum_{\mu=1}^p \tilde{\xi}_{i\mu}\tilde{\xi}_{j\mu}$.

Let us recall a general mathematical framework for studying the convergence of random empirical measures. 
\begin{definition}\label{def:1}
    Let $\mathcal{P}(\R)$ denote the space of probability measures on $\R$ endowed with the topology of weak convergence. We say that a sequence of random probability measures $\{\rho_N\}_{N=1}^\infty$ in $\mathcal{P}(\R)$ converges in probability to a deterministic probability measure $\rho \in \mathcal{P}(\R)$ if, for all continuous and bounded functional $F: \mathcal{P}(\R) \to\R$,
    \begin{equation*}
        \E[F(\rho_N)] \to \E[F(\rho)], \quad \text{as }N\to\infty.
    \end{equation*}
\end{definition}
In the field the interacting particle systems, It is well known that to show the convergence defined as in Definition~\ref{def:1}, it suffices to show that for all continuous and bounded functions  $f:\R \to \R$,
\begin{equation*}
    \E\left[\left|\int f(z)\rho_N(\rd z) - \int f(z)\rho(\rd z)\right|\right] \to 0,\quad \text{as }N\to\infty; 
\end{equation*}
see for example \cite[Proposition~4.2]{Mel06}. (The functions $f$ are usually called ``test functions''.)

We distinguish two cases: first, the case where $p\in\mathbb{N}_+$ is fixed as $N\to\infty$ (Theorem~\ref{theorem:1}); second , the case where $p=\alpha N$, for $\alpha>0$, as $N\to\infty$ (Theorem~\ref{theorem:2}). To prove both theorems, we will use the following lemma which does not depend on the scaling of $p$ as $N\to\infty$.

\begin{lemma}\label{lemma:1}
    For any $N>1$ and $p>0$, let $\{\zeta_{i,\mu}\}_{1\leq i \leq N, 1\leq \mu \leq p}$ be a collection of $i.i.d.$ random variables. For all bounded functions $f:\R\to\R$,
    \begin{equation}\label{eq:lemma_variance_bound}
        \mathrm{Var}\left(\frac{2}{N(N-1)} \sum_{1\leq i < j \leq N}f\left(\frac{1}{p}\sum_{\mu=1}^p \zeta_{i\mu}\zeta_{j\mu}\right)\right) \leq \frac{2(2N-3)}{N(N-1)}\|f\|_\infty^2
    \end{equation}
In particular, the variance bound~\eqref{eq:lemma_variance_bound} does not depend on $p$ and it vanishes as $N\to\infty$.
\end{lemma}

\begin{proof}
    We use the shorthand $f(i,j):= f\left(\frac{1}{p}\sum_{\mu=1}^p \zeta_{i\mu}\zeta_{j\mu}\right)$.
    \begin{align}
        \mathrm{Var}\left(\frac{2}{N(N-1)} \sum_{1\leq i < j \leq N}f(i,j)\right) 
        &= \frac{4}{N^2(N-1)^2}\left\{\E\left[\left(\sum_{1\leq i < j \leq N}f(i,j)\right)^2\right] - \left(\E\left[\sum_{1\leq i < j \leq N}f(i,j)\right]\right)^2\right\} \nonumber\\
        &= \frac{4}{N^2(N-1)^2}\sum_{1\leq i<j\leq N} \sum_{1\leq k<l\leq N}\E[f(i,j)f(k,l)] - \E[f(i,j)]\E[f(k,l)]. \label{eq:double_sum}
    \end{align}
    If $\{i,j\}\cap\{k,l\}=\varnothing$, $f(i,j)$ and $f(k,l)$ are independent and $\E[f(i,j)f(k,l)] - \E[f(i,j)]\E[f(k,l)]=0$. If $\{i,j\}\cap\{k,l\}\neq\varnothing$, $\E[f(i,j)f(k,l)] - \E[f(i,j)]\E[f(k,l)]$ can be upper-bounded by $\|f\|_\infty^2$. Since, in the double sum~\eqref{eq:double_sum}, $\{i,j\}\cap\{k,l\}=\varnothing$ for $N(N-1)/2 \times (N-2)(N-3)/2$ of the summands, we know that $\{i,j\}\cap\{k,l\}\neq\varnothing$ for 
    \begin{equation*}
        \frac{N(N-1)}{2}\frac{N(N-1)}{2} - \frac{N(N-1)}{2} \frac{(N-2)(N-3)}{2} = \frac{N(N-1)}{2}(2N - 3)
    \end{equation*}
    of the summands. Hence, we get that
    \begin{multline*}
        \frac{4}{N^2(N-1)^2}\sum_{1\leq i<j\leq N} \sum_{1\leq k<l\leq N}\E[f(i,j)f(k,l)] - \E[f(i,j)]\E[f(k,l)] \\
        \leq \frac{4}{N^2(N-1)^2}\frac{N(N-1)}{2}(2N-3)\|f\|_\infty^2 = \frac{2(2N-3)}{N(N-1)}\|f\|_\infty^2. 
    \end{multline*}
\end{proof}

\begin{theorem}\label{theorem:1}
For any fixed $p>1$, the sequence of empirical measures  
\begin{equation*}
    \rho_N(\rd z) := \frac{2}{N(N-1)} \sum_{1\leq i < j \leq N}\delta_{\frac{1}{p}\sum_{\mu=1}^p \tilde{\xi}_{i\mu}\tilde{\xi}_{j\mu}}(\rd z)\qquad \forall \,N>1,
\end{equation*}
converges in probability (according to Definition~\ref{def:1}), as $N\to\infty$, to the deterministic probability measure $\rho\in\mathcal{P}(\R)$ given by
\begin{equation*}
    \rho(\rd z) := \mathbbm{1}_{[-1,1]}(z)\frac{\Gamma(p/2)}{\sqrt{\pi}\,\Gamma((p-1)/2)}(1-z^2)^{\frac{p-3}{2}}\rd z.
\end{equation*}
\end{theorem}
\begin{proof}
As mentioned above, to show the convergence of $\rho_N$ to $\rho$ in probability, it suffices to show that for all continuous and bounded functions $f:\R\to\R$, 
\begin{equation*}
    \E\left[\left|\int f(z)\rho_N(\rd z) - \int f(z)\rho(\rd z)\right|\right] \to 0,\quad \text{as }N\to\infty.
\end{equation*}
%By Cauchy-Schwarz inequality, 
%\begin{align*}
%    &\E\left[\left|\int f(z)\nu_N(\rd z) - \E\left[\int f(z)\nu_N(\rd z)\right]\right|\right] \\
%    &\qquad = \E\left[\left|\int f(z)\nu_N(\rd z) - \E\left[\int f(z)\nu_N(\rd z)\right]\right|\cdot \mathbbm{1}\right]\\
%    &\qquad \leq \E\left[\left(\int f(z)\nu_N(\rd z) - \E\left[\int f(z)\nu_N(\rd z)\right]\right)^2\right]^{1/2}\underbrace{\E\left[\mathbbm{1}^2\right]^{1/2}}_{=1} \\
%    &\qquad= \left[\mathrm{Var}\left(\int f(z)\nu_N(\rd z)\right) \right]^{1/2}
%\end{align*}
By Jensen's inequality \cite[Theorem~5.1 p.~132]{Gut06} and Lemma~\ref{lemma:1}, 
\begin{equation*}
    \E\left[\left|\int f(z)\rho_N(\rd z) - \E\left[\int f(z)\rho_N(\rd z)\right]\right|\right] \leq \left[\mathrm{Var}\left(\int f(z)\rho_N(\rd z)\right) \right]^{1/2}\xrightarrow[N\to\infty]{}0.
\end{equation*}
Hence, it only remains to show that, for all $N>1$,
\begin{equation}\label{eq:thm_1_expectation}
    \E\left[\int f(z)\rho_N(\rd z)\right] = \int f(z)\rho(\rd z) = \int f(z)\frac{\Gamma(p/2)}{\sqrt{\pi}\,\Gamma((p-1)/2)}(1-z^2)^{\frac{p-3}{2}}\rd z.
\end{equation}
Indeed, we have
\begin{align}
    \E\left[\int f(z)\rho_N(\rd z)\right] &= \E\left[\frac{2}{N(N-1)}\sum_{1\leq i < j \leq N}f\left(\frac{1}{p}\sum_{\mu=1}^p \tilde{\xi}_{i\mu}\tilde{\xi}_{j\mu}\right)\right] = \E\left[f\left(\frac{1}{p}\sum_{\mu=1}^p \tilde{\xi}_{1\mu}\tilde{\xi}_{2\mu}\right)\right] \nonumber\\
    &= \E\left[\E\left[f\left(\frac{1}{p}\sum_{\mu=1}^p \tilde{\xi}_{1\mu}\tilde{\xi}_{2\mu}\right)\,\Bigg|\,\tilde{\xi}_{1,1},\dots,\tilde{\xi}_{1,p}\right]\right]. \label{eq:conditional_expectation}
\end{align}
By definition, the random vector $\frac{1}{\sqrt{p}}(\tilde{\xi}_{1,1},\dots,\tilde{\xi}_{1,p})^T$ lies on the $(p-1)$-sphere of unit radius, $\mathbb{S}^{p-1}$, and, by symmetry, the conditional expectation in \eqref{eq:conditional_expectation} is constant for all $\frac{1}{\sqrt{p}}(\tilde{\xi}_{1,1},\dots,\tilde{\xi}_{1,p})^T \in \mathbb{S}^{p-1}$. Since the vector $\frac{1}{\sqrt{p}}(\tilde{\xi}_{2,1},\dots,\tilde{\xi}_{2,p})^T$ is uniformly distributed on $\mathbb{S}^{p-1}$, we have, by the Funk-Hecke formula (see \cite[Theorem~1.2.9]{DaiXu13} for a reference),
\begin{equation*}
    \E\left[\E\left[f\left(\frac{1}{p}\sum_{\mu=1}^p \tilde{\xi}_{1\mu}\tilde{\xi}_{2\mu}\right)\,\Bigg|\,\tilde{\xi}_{1,1},\dots,\tilde{\xi}_{1,p}\right]\right] = \int_{-1}^1 f(z)\frac{\Gamma(p/2)}{\sqrt{\pi}\,\Gamma((p-1)/2)}(1-z^2)^{\frac{p-3}{2}}\rd z,
\end{equation*}
which concludes the proof of \eqref{eq:thm_1_expectation}.
\end{proof}
\begin{corollary}
    If $\theta_1, \dots, \theta_N$ are $i.i.d.$ and uniformly distributed on the interval $[0,2\pi[$, the sequence of empirical measures
    \begin{equation*}
        \rho_N(\rd z) := \frac{2}{N(N-1)} \sum_{1\leq i < j \leq N}\delta_{\cos(\theta_i - \theta_j)}(\rd z)
    \end{equation*}
    converges in probability (according to Definition~\ref{def:1}), as $N\to\infty$, to the deterministic probability measure $\rho\in\mathcal{P}(\R)$ given by
    \begin{equation*}
        \rho(\rd z) := \mathbbm{1}_{[-1,1]}(z)\frac{1}{\pi}\frac{1}{\sqrt{1 - z^2}}\rd z. 
    \end{equation*}
\end{corollary}
\begin{proof}
    Notice that the collection of random variables $\{\cos(\theta_i - \theta_j)\}_{i<j}$ has the same joint law as $\{\frac{1}{2}(\tilde{\xi}_{i,1}\tilde{\xi}_{j,1} + \tilde{\xi}_{i,2}\tilde{\xi}_{j,2})\}_{i<j}$. Thus, the result follows from the Theorem~\ref{theorem:1} with $p=2$.
\end{proof}

\begin{theorem}\label{theorem:2}
If $p=\alpha N$ for some fixed $\alpha>0$, the sequence of empirical measures
\begin{equation*}
    \widehat{\rho_N}(\rd z) := \frac{2}{N(N-1)} \sum_{1\leq i < j \leq N}\delta_{\frac{1}{\sqrt{p}}\sum_{\mu=1}^p \tilde{\xi}_{i\mu}\tilde{\xi}_{j\mu}}(\rd z)\qquad \forall \,N>1,
\end{equation*}
converge in probability (according to Definition~\ref{def:1}), as $N\to\infty$, to the deterministic probability measure $\rho\in\mathcal{P}(\R)$ given by
\begin{equation*}
    \rho(\rd z) := \frac{1}{\sqrt{2\pi}}e^{-z^2/2}\rd z.
\end{equation*}
\end{theorem}
\begin{proof}
As in the proof of Theorem~\ref{theorem:1}, to show that the convergence in probability of $\rho_N$ to $\rho$, it suffices to show that for all continuous and bounded function $f:\R\to\R$, 
\begin{equation*}
    \E\left[\left|\int f(z)\widehat{\rho_N}(\rd z) - \int f(z)\rho(\rd z)\right|\right] \to 0,\quad \text{as }N\to\infty.
\end{equation*}
By triangular inequality, 
\begin{multline}\label{eq:theorem_2_triangular}
\E\left[\left|\int f(z)\widehat{\rho_N}(\rd z) - \int f(z)\rho(\rd z)\right|\right] \\
\leq \E\left[\left|\int f(z)\widehat{\rho_N}(\rd z) - \E\left[\int f(z)\widehat{\rho_N}(\rd z)\right]\right|\right] + \left|\E\left[\int f(z)\widehat{\rho_N}(\rd z)\right] - \int f(z)\rho(\rd z)\right|.
\end{multline}
By Jensen's inequality and Lemma~\ref{lemma:1}, we know the first term on the right-hand side of \eqref{eq:theorem_2_triangular} vanishes as $N\to\infty$ and it only remains to show that 
\begin{equation}\label{eq:theorem_2_expectation}
    \E\left[\int f(z)\widehat{\rho_N}(\rd z)\right] \to \int f(z)\rho(\rd z), \quad\text{ as }N\to\infty.
\end{equation}
Since
\begin{equation*}
    \E\left[\int f(z)\widehat{\rho_N}(\rd z)\right] = \E\left[f\left(\frac{1}{\sqrt{p}}\sum_{\mu=1}^p \tilde{\xi}_{1\mu}\tilde{\xi}_{2\mu}\right)\right],
\end{equation*}
showing \eqref{eq:theorem_2_expectation} is equivalent to showing
\begin{equation*}
    \E\left[f\left(\frac{1}{\sqrt{p}}\sum_{\mu=1}^p \tilde{\xi}_{1\mu}\tilde{\xi}_{2\mu}\right)\right] \to \int f(z)\rho(\rd z), \quad\text{ as }N\to\infty,
\end{equation*}
which in turn is equivalent to showing that the random variables $\widetilde{X}_N := \frac{1}{\sqrt{p}}\sum_{\mu=1}^p \tilde{\xi}_{1\mu}\tilde{\xi}_{2\mu}$ converge in law to a standard normal variable $X\sim\mathcal{N}(0,1)$ as $N\to\infty$. We recall that $p=\alpha N$. As $N\to\infty$, we have, on the one hand, that $\sqrt{\frac{1}{p}\sum_{\mu=1}^p\xi_{i\mu}^2}$ converges in probability to $1$ by the weak law of large numbers and the continuous mapping theorem \cite[Theorem~10.3 p.~245]{Gut06}, and on the other hand, $X_N := \frac{1}{\sqrt{p}}\sum_{\mu=1}^p \xi_{1\mu}\xi_{2\mu}$ converges in law to $X$ by the central limit theorem. Hence, $\widetilde{X}_N$ converges in law to $X$ by Slutsky's theorem \cite[Theorem~11.4 p.~249]{Gut06}. This proves \eqref{eq:theorem_2_expectation} and concludes the proof.
\end{proof}

\ensureoddstart
\section{From low-rank SNNs to neural field equations}

For any number of neurons $N$ and any number of patterns $p$, let $\bxi$ be a random $N \times p$-matrix with \textit{i.i.d.}, zero-mean, unit-variance, normally distributed entries $\{\xi_{i\mu}\}_{i,\mu}$. We choose a connectivity matrix given by
\begin{equation*}
    J_{ij} := \frac{1}{cN}\sum_{\mu=1}^p \xi_{i\mu}\left(\phi(\xi_{j\mu}) - a\right) \quad \text{for all }i\neq j,
\end{equation*}
and $J_{ii} := 0$ for all $i$. As in the main text, $a:=\int_{-\infty}^\infty \mathcal{D}z\; \phi(z)$ and $c:=\int_{-\infty}^\infty \mathcal{D}z\; (\phi(z) - a)^2$, where $\mathcal{D}z$ is the standard Gaussian measure. This type of connectivity matrix was introduced by Brunel and Pereira \cite{PerBru18} in the context of Hopfield networks.

When the number of patterns $p$ is kept constant as the number of neurons $N$ tends to infinity, we say that the connectivity matrix $\J = \{J_{ij}\}_{\substack{1\leq i \leq N \\ 1\leq j \leq N}}$ is low-rank since its rank remains finite as $N\to\infty$. Although large, random, low-rank networks of neurons \cite{MasOst18,SchDub20} are usually not presented as spatially structured networks, they do have an implicit spatial structure. By identifying this spatial structure, we can derive exact neural field equations describing the dynamics of large networks, as we show here for the connectivity matrix $\J$.

For simplicity, let us first consider an SNN of $N$ neurons without external input. The $N$ neurons are assigned to $N$ points in $\R^p$ corresponding to the rows of $\bxi$, i.e., neuron $i$ is assigned to the point $\bxi_i = (\xi_{i1}, \dots, \xi_{ip})$. Writing $u(t,\bxi_i)= h_i(t)$ for all $i=1, \dots,N$, we can interpret $u(t,\bxi_i)$ as being the potential of a neuron at location $\bxi_i\in\R^p$ at time $t$. Without loss of generality, let us consider the dynamics of neuron $i=1$, which can be rewritten
\begin{align*}
    \tau \frac{\rd}{\rd t}u(t,\bxi_1) &= -u(t,\bxi_1) + \sum_{j=2}^N\underbrace{\frac{1}{cN}\sum_{\mu=1}^p \xi_{1\mu}\left(\phi(\xi_{j\mu}) - a\right)}_{=J_{1j}}s_j(t), \\
    &= -u(t,\bxi_1) + \frac{1}{c}\sum_{\mu=1}^p \xi_{1\mu}\frac{1}{N}\sum_{j=2}^N\left(\phi(\xi_{j\mu}) - a\right)s_j(t),
\end{align*}
where $s_j(t)$ has instantaneous firing rate $\phi(u(t,\bxi_j))$. By spatial averaging \cite{Ger95,GerKis14}, we have
\begin{align}
    \lim_{N\to\infty} \frac{1}{N}\sum_{j=2}^N\left(\phi(\xi_{j\mu}) - a\right)s_j(t) &= \lim_{N\to\infty} \frac{1}{N}\sum_{j=2}^N\left(\phi(\xi_{j\mu}) - a\right)\phi(u(t,\bxi_j))\nonumber \\
    &= \int_{\R^p} \underbrace{\lim_{N\to\infty}\frac{1}{N}\sum_{j=2}^N\delta_{\bxi_j}(d\bz)}_{=\mathcal{D}\bz}\left(\phi(z_\mu) - a\right)\phi(u(t,\bz))\nonumber \\
    &= \int_{\R^p}\mathcal{D}\bz\,\left(\phi(z_\mu) - a\right)\phi(u(t,\bz)), \label{eq:emp_to_Dz}
\end{align}
where $\delta_{\bxi_j}(\rd\bz)$ denotes the Dirac measure centered on the point $\bxi_j$ and $\mathcal{D}\bz$ denotes the $p$-dimensional standard Gaussian measure
\begin{equation*}
    \mathcal{D}\bz = dz_1 \dots dz_p \frac{1}{(2\pi)^{p/2}}e^{-\|\bz\|_2^2}.
\end{equation*}
Intuitively, the equalities~\eqref{eq:emp_to_Dz} simply reflect the fact that as $N\to\infty$, the empirical measure $\frac{1}{N}\sum_{i=1}^N\delta_{\bxi_i}$ on $\R^p$, which summarizes the locations $\bxi_1, \bxi_2, \dots, \bxi_N$ of neurons in $\R^p$, converges to the standard Gaussian measure $\mathcal{D}\bz$ on $\R^p$ (since the $p$-dimensional random vectors $\bxi_1, \bxi_2, \dots, \bxi_N$ are \textit{i.i.d.} with distribution $\mathcal{D}\bz$), turning the summation over neurons into an integration over a Gaussian measure. Therefore, since the spatial distribution of neurons in $\R^p$ converges to a continuous Gaussian distribution as $N\to\infty$, we find that the continuous field $u(t,\by)$, with $\by\in\R^p$, solves the integro-differential equation
\begin{equation*}
    \tau \frac{\partial}{\partial t}u(t,\by) = - u(t,\by) + \frac{1}{c}\sum_{\mu=1}^p y_\mu \int_{\R^p}\mathcal{D}\bz\,\left(\phi(z_\mu) - a\right)\phi(u(t,\bz)), \qquad \forall\by\in\R^p.
\end{equation*}
Defining the connectivity kernel $w:\R^p \to \R^p$
\begin{equation}
    w(\by,\bz) := \frac{1}{c}\sum_{\mu=1}^p y_\mu \left(\phi(z_\mu) - a\right),
\end{equation}
we get a neural field equation (see \cite{Bre11} and Eq.~6.129 in \cite[p.~244]{GerKis02}):
\begin{equation}\label{eq:NFE}
    \tau \frac{\partial}{\partial t}u(t,\by) = - u(t,\by) + \int_{\R^p}\mathcal{D}\bz\,w(\by,\bz)\phi(u(t,\bz)), \qquad \forall\by\in\R^p.
\end{equation}
While the arguments presented here are informal, the neural field equation is the exact mean-field limit of the low-rank SNN: using the embedding of the low-rank SNN in $\R^p$, the convergence of the SNN to the neural field equation~\eqref{eq:NFE} is guaranteed by rigorous results for spatially structured SNNs \cite{CheDua19}.

Of course, by the same arguments, the corresponding low-rank RNN also converges to the neural field equation~\eqref{eq:NFE}. {\color{black}While the convergence of low-rank SNNs/RNNs to neural fields equations is, to our knowledge, a new result, the idea of using the rows of the $N\times p$ pattern matrix $\bxi$ to define an embedding of the neurons in $\R^p$ is not new: it has been used to study the \textit{equilibrium} statistical mechanics of certain spin-based neural networks \cite{HemGre86,HemGre88}.}

In the case where half of the neurons receive an external input $\frac{\sigma}{\sqrt{p}}\sum_{\mu=1}^p \xi_{i\mu}\eta_\mu(t)$, as defined in the main text, we can describe the population dynamics with two neural field equations. Following the same steps as above, we get that for each neuron $i$ receiving external input, $\lim_{N\to\infty}h^{\mathrm{in}}_i(t) = \lim_{N\to\infty}x^{\mathrm{in}}_i(t) = u^{\mathrm{in}}(t,\bxi_i)$, and for all neuron $i'$ receiving no external input but only recurrent input, $\lim_{N\to\infty}h^{\mathrm{rec}}_{i'}(t) = \lim_{N\to\infty}x^{\mathrm{rec}}_{i'}(t) = u^{\mathrm{rec}}(t,\bxi_{i'})$, where the fields $u^{\mathrm{in}}$ and $u^{\mathrm{rec}}$ are the solutions to the system
\begin{align*}
    \tau \frac{\partial}{\partial t}u^{\mathrm{in}}(t,\by) &= - u^{\mathrm{in}}(t,\by) + \frac{1}{2}\int_{\R^p}\mathcal{D}\bz\,w(\by,\bz)\phi(u^{\mathrm{in}}(t,\bz))+\frac{1}{2}\int_{\R^p}\mathcal{D}\bz\,w(\by,\bz)\phi(u^{\mathrm{rec}}(t,\bz)) + \frac{\sigma}{\sqrt{p}}\sum_{\mu=1}^p y_\mu\eta_\mu(t), \\
    \tau \frac{\partial}{\partial t}u^{\mathrm{rec}}(t,\by) &= - u^{\mathrm{in}}(t,\by) + \frac{1}{2}\int_{\R^p}\mathcal{D}\bz\,w(\by,\bz)\phi(u^{\mathrm{in}}(t,\bz))+\frac{1}{2}\int_{\R^p}\mathcal{D}\bz\,w(\by,\bz)\phi(u^{\mathrm{rec}}(t,\bz)),
\end{align*}
where the $\{\eta_\mu(t)\}_{\mu=1}^p$ are the formal derivatives of independent Wiener processes (or standard Brownian motions) $B_1(t), \dots, B_p(t)$, i.e. $\eta_\mu(t) = \rd B_\mu(t)/\rd t$.
\ensureoddstart

\section{Bound on the probability of a duplicate}
Let us assume that the correlations $\{C_{ij}\}_{1\leq i,j\leq N}$ are given by \eqref{eq:corr_def_appendix} {\color{black}(Approximation~(9) in the main text)}.% (which is equivalent to using the approximation~\eqref{eq:correl} in the main text).
\begin{lemma}
    For any $N,p>1$ and for any $0<\varepsilon<1$, we have the bound
    \begin{equation*}
        \mathbb{P}(\exists\, i \neq j \text{ such that } |C_{ij}|\geq \varepsilon) \leq \frac{N(N-1)}{2}\frac{(1-\varepsilon^2)^{\frac{p-1}{2}}}{\sqrt{\pi}}.
    \end{equation*}
\end{lemma}
\begin{proof}
By union bound, we have
\begin{equation*}
     \mathbb{P}(\exists\, i \neq j \text{ such that } |C_{ij}|\geq \varepsilon) \leq \frac{N(N-1)}{2} \mathbb{P}(|C_{12}|\geq \varepsilon).
\end{equation*}
Then, we use the fact that $C_{12}$ follows a Gegenbauer distribution with parameter $p$, whence
\begin{equation*}
    \mathbb{P}(|C_{12}|\geq \varepsilon) = 2\int_\varepsilon^1 \frac{\Gamma(p/2)}{\sqrt{\pi}\,\Gamma((p-1)/2)}(1-z^2)^{\frac{p-3}{2}}\rd z.
\end{equation*}
First, we notice that
\begin{equation*}
    \frac{\Gamma(p/2)}{\Gamma((p-1)/2)} \leq \frac{\Gamma((p+1)/2)}{\Gamma((p-1)/2)} = \frac{p-1}{2};
\end{equation*}
second, using the change of variable $y^2 = 1 - z^2$, 
\begin{equation*}
    \int_\varepsilon^1 (1-z^2)^{\frac{p-3}{2}}\rd z = \int_0^{\sqrt{1-\varepsilon^2}} \left[y^2\right]^{\frac{p-3}{2}}\frac{y}{\sqrt{1-y^2}}\rd y \leq \int_0^{\sqrt{1-\varepsilon^2}} y^{p-2}\rd y = \frac{(1-\varepsilon^2)^{\frac{p-1}{2}}}{p-1}.
\end{equation*}
Hence, we have shown that 
\begin{equation*}
    \mathbb{P}(|C_{12}|\geq \varepsilon) \leq \frac{(1-\varepsilon^2)^{\frac{p-1}{2}}}{\sqrt{\pi}},
\end{equation*}
which concludes the proof.
\end{proof}

\ensureoddstart
\section{Concentration of the $\ell^2$ norms of the incoming weights}
We recall that the $\ell^2$ norm of the incoming weights to neuron $i$ is $\|\J_i\|_2 = \sqrt{\sum_{j=1}^N J_{ij}^2}$ and that the weights are given by $J_{ij} := \frac{1}{cN}\sum_{\mu=1}^p \xi_{i\mu}\left(\phi(\xi_{j\mu}) - a\right)$, for all $i\neq j$, and $J_{ii} := 0$, where $a=\int_{-\infty}^\infty \mathcal{D}z\; \phi(z)$ and $c=\int_{-\infty}^\infty \mathcal{D}z\; (\phi(z) - a)^2$ are constants ($\mathcal{D}z$ denotes the standard Gaussian measure).
\begin{theorem}\label{theorem:weight_concentration}
    When $p = \alpha N$ for some fixed $\alpha>0$, the $\ell^2$ norm of the incoming weights of a typical neuron $i$ converges in probability to $\sqrt{\alpha/c}$ as $N\to\infty$, i.e. 
    \begin{equation*}
        \|\J_{i}\|_2 \xrightarrow[N\to\infty]{\mathbb{P}} \sqrt{\frac{\alpha}{c}}.
    \end{equation*}
\end{theorem}
\begin{proof}
    Without loss of generality, we can take $i=1$. The proof can be decomposed in two steps. First (Step 1), we will show that
    \begin{equation}\label{eq:exp_l2}
        \lim_{N\to\infty}\E\left[\|\J_1\|_2^2\right] = \frac{\alpha}{c},
    \end{equation}
    then (Step 2), we will show that 
    \begin{equation}\label{eq:var_l2}
        \Var\left(\|\J_1\|_2^2\right) = \frac{2\alpha(1 + \alpha) }{c^2 N}+ \mathcal{O}\left(\frac{1}{N^2}\right).
    \end{equation}
    If \eqref{eq:exp_l2} and \eqref{eq:var_l2} are verified, $\|\J_1\|_2^2$ converges in probability to $\alpha/c$ and, by the continuous mapping theorem, $\|\J_1\|_2$ converges in probability to $\sqrt{\alpha/c}$.

\underline{Step 1:}

\begin{align*}
    \E\left[\|\J_1\|_2^2\right] &= \E\left[\sum_{j=1}^N J^2_{1j}\right] = \sum_{j=1}^N \E\left[ J^2_{1j}\right] = \frac{1}{c^2N^2}\sum_{j=2}^N \E\left[ \left(\sum_{\mu=1}^p \xi_{1\mu}\left(\phi(\xi_{j\mu}) - a\right)\right)^2\right] \\
    %&= \frac{1}{c^2N^2}\sum_{\substack{j = 1 \\ j\neq i}}^N \E\left[ \sum_{\mu=1}^p \xi_{i\mu}^2\left(\phi(\xi_{j\mu}) - a\right)^2 + \sum_{\mu\neq \nu}\xi_{i\mu}\left(\phi(\xi_{j\mu}) - a\right)\xi_{i\nu}\left(\phi(\xi_{j\nu}) - a\right)\right] \\
    &\stackrel{(\ast)}{=} \frac{1}{c^2N^2}\sum_{j=2}^N \E\left[ \sum_{\mu=1}^p \xi_{1\mu}^2\left(\phi(\xi_{j\mu}) - a\right)^2\right] = \frac{1}{c^2N^2}\sum_{j=2}^N \sum_{\mu=1}^p \E\left[  \xi_{1\mu}^2\left(\phi(\xi_{j\mu}) - a\right)^2\right] \\
    %&= \frac{1}{c^2 N^2}(N-1)p\E\left[\xi_{11}^2\left(\phi(\xi_{21})-a\right)^2\right] \\
    &= \frac{1}{c^2 N^2}(N-1)\alpha N c \xrightarrow[N\to\infty]{}\frac{\alpha}{c}
\end{align*}
For the equality $(\ast)$, we used the fact that the cross terms are null since the columns of the random matrix $\bxi$, i.e. the patterns, are independent and when $\mu\neq\nu$,
\begin{equation*}
    \E\left[\xi_{1\mu}\left(\phi(\xi_{j\mu})-a\right)\xi_{1\nu}\left(\phi(\xi_{j\nu})-a\right)\right] = \E\left[\xi_{1\mu}\left(\phi(\xi_{j\mu})-a\right)\right]\E_{\bxi}\left[\xi_{1\nu}\left(\phi(\xi_{j\nu})-a\right)\right] = 0.
\end{equation*}

\underline{Step 2:}
\begin{equation}\label{eq:variance}
    \Var\left(\|\J_1\|_2^2\right) = \Var\left(\sum_{j=1}^N J_{1j}^2\right) = \E\left[\left(\sum_{j=1}^N J^2_{1j}\right)^2\right] - \E\left[\sum_{j=1}^N J_{1j}^2\right]^2. 
\end{equation}
By the computation in Step 1 above, we have 
\begin{equation}\label{eq:E2}
    \E\left[\sum_{j=1}^N J_{1j}^2\right]^2 = \frac{1}{c^2}\frac{(N-1)^2p^2}{N^4}.
\end{equation}
The second moment can be decomposed as
\begin{equation}
    \E\left[\left(\sum_{j=1}^N J^2_{1j}\right)^2\right] = \E\left[\sum_{j=1}^N J_{1j}^4 + \sum_{j=1}^N\sum_{k\neq j}J_{1j}^2 J_{1k}^2\right] = (N-1)\E\left[J_{12}^4\right] + (N-1)(N-2)\E\left[J_{12}^2 J_{13}^2\right], \label{eq:second_moment}
\end{equation}
and we compute the terms $\E\left[J_{12}^4\right]$ and $\E\left[J_{12}^2 J_{13}^2\right]$ separately. 
\begin{align*}
    \E\left[J_{12}^4\right] &= \frac{1}{c^4N^4}\E\left[\left(\sum_{\mu=1}^p\xi_{1\mu}\left(\phi(\xi_{2\mu}) - a\right)\right)^4\right] \nonumber\\
    &=\frac{1}{c^4N^4}\sum_{\mu=1}^p \E\left[\xi_{1\mu}^4\left(\phi(\xi_{2\mu}) - a\right)^4\right] + \frac{1}{c^4N^4}\sum_{1\leq\mu < \nu \leq p}{4\choose 2}\E\left[\xi_{1\mu}^2\left(\phi(\xi_{2\mu}) - a\right)^2\xi_{1\nu}^2\left(\phi(\xi_{2\nu}) - a\right)^2\right]
\end{align*}
where for the second equality, we used the fact that, in the multinomial expansion of
\begin{equation*}
    \E\left[\left(\sum_{\mu=1}^p\xi_{1\mu}\left(\phi(\xi_{2\mu}) - a\right)\right)^4\right],
\end{equation*} 
all terms with at least one odd power are null. Hence, we get that
\begin{equation}
    \E\left[J_{12}^4\right] 
    = \mathcal{O}\left(\frac{1}{N^3}\right) + \frac{1}{c^4 N^4}\frac{p(p-1)}{2}6c^2 =\frac{3}{c^2 N^4}p(p-1) + \mathcal{O}\left(\frac{1}{N^3}\right). \label{eq:J_12}
\end{equation}

Similarly,
\begin{align}
    \E\left[J_{12}^2 J_{13}^2\right] &= \frac{1}{c^4 N^4}\E\left[\left(\sum_{\mu=1}^p\xi_{1\mu}\left(\phi(\xi_{2\mu}) - a\right)\right)^2\left(\sum_{\mu=1}^p\xi_{1\mu}\left(\phi(\xi_{3\mu}) - a\right)\right)^2\right] \nonumber\\
    &= \frac{1}{c^4 N^4}\sum_{\mu=1}^p \E\left[\xi_{1\mu}^4\left(\phi(\xi_{2\mu})-a\right)^2\left(\phi(\xi_{3\mu})-a\right)^2\right] +\frac{2}{c^4 N^4}\sum_{1\leq\mu<\nu\leq p}\E\left[\xi_{1\mu}^2\left(\phi(\xi_{2\mu})-a\right)^2\xi_{1\nu}^2\left(\phi(\xi_{3\nu})-a\right)^2\right] \nonumber\\
    %&= \frac{1}{c^4 N^4}p3c^2 + \frac{2}{c^4 N^4}\frac{p(p-1)}{2}c^2 \nonumber\\
    &= \frac{3}{c^2 N^4}p + \frac{1}{c^2 N^4}p(p-1). \label{eq:J_12J13}
\end{align}
From Eqs.~\eqref{eq:variance}, \eqref{eq:E2}, \eqref{eq:second_moment}, \eqref{eq:J_12} and \eqref{eq:J_12J13}, we get
\begin{align*}
    \Var\left(\|\J_1\|_2^2\right) &= \E\left[\left(\sum_{j=1}^N J^2_{ij}\right)^2\right] - \E\left[\sum_{j=1}^N J_{ij}^2\right]^2 \\
    &= (N-1)\E\left[J_{12}^4\right] + (N-1)(N-2)\E\left[J_{12}^2 J_{13}^2\right] - \frac{1}{c^2 N^4}(N-1)^2p^2 \\
    &= \frac{3}{c^2}\frac{(N-1)p(p-1)}{N^4}  + \frac{3}{c^2}\frac{(N-1)(N-2)p}{N^4} + \frac{1}{c^2}\frac{(N-1)(N-2)p(p-1)}{N^4}  - \frac{1}{c^2}\frac{(N-1)^2p^2}{N^4} + \mathcal{O}\left(\frac{1}{N^2}\right).
\end{align*}
Using the fact that
\begin{align*}
    (N-1)^2p^2 &= (N-1)(N-2)p^2 + (N-1)p^2 \\
    &= (N-1)(N-2)p(p-1) + (N-1)(N-2)p + (N-1)p(p-1) + (N-1)p,
\end{align*}
and after some rearrangement of terms, we obtain, for large $N$,
\begin{align*}
    \Var\left(\|\J_1\|_2^2\right) 
    %&= \frac{3}{c^2}\frac{(N-1)p(p-1)}{N^4} - \frac{1}{c^2}\frac{(N-1)p(p-1)}{N^4}\\
    %&\qquad + \frac{3}{c^2}\frac{(N-1)(N-2)p}{N^4} -   \frac{1}{c^2}\frac{(N-1)(N-2)p}{N^4}\\
    %&\qquad + \frac{1}{c^2}\frac{(N-1)(N-2)p(p-1)}{N^4} - \frac{1}{c^2}\frac{(N-1)(N-2)p(p-1)}{N^4} \\
    %&\qquad + \mathcal{O}\left(\frac{1}{N^2}\right) \\
    %&= \frac{2}{c^2}\frac{(N-1)p(p-1)}{N^4} + \frac{2}{c^2}\frac{(N-1)(N-2)p}{N^4} + \mathcal{O}\left(\frac{1}{N^2}\right) \\
    &= \frac{2\alpha(1 + \alpha) }{c^2 N}+ \mathcal{O}\left(\frac{1}{N^2}\right).
\end{align*}
\end{proof}

\ensureoddstart
\section{Simplified feedforward model}\label{sec:feedforward}
\begin{figure*}[ht]
    \includegraphics[width=\textwidth]{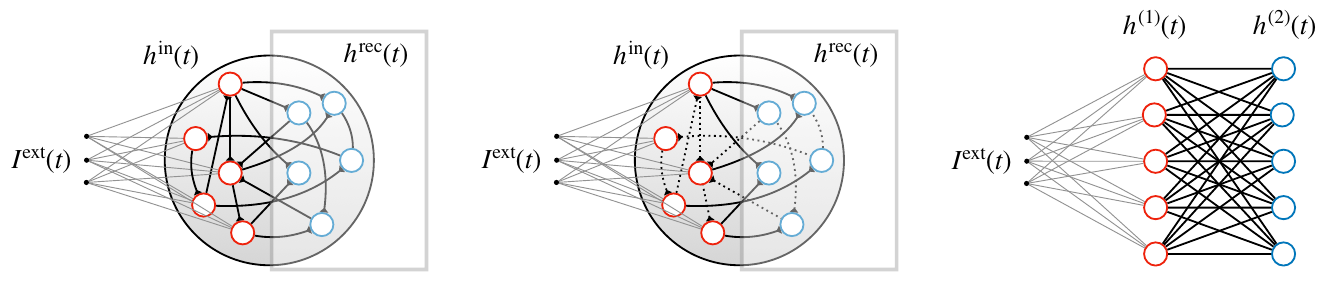}
    \caption{\label{fig:ff}Feedforward model as an simplification of the input-driven model.}
\end{figure*}

To understand mathematically why large SNNs converge to large RNNs as the load $\alpha\to0$, we consider a simplified feedforward model. This simplified model is obtained by keeping network connections from `in' to `rec' neurons and removing all the other connections. After this pruning procedure, we are left with two-layer feedforward networks where the `in' neurons make up the first layer, $(h^{\mathrm{in}}_i(t), x^{\mathrm{in}}_i(t)) \mapsto (h^{(1)}_i(t), x^{(1)}_i(t))$, and the `rec' neurons make up the second layer, $(h^{\mathrm{rec}}_i(t), x^{\mathrm{rec}}_i(t)) \mapsto (h^{(2)}_i(t), x^{(2)}_i(t))$ (Fig.~\ref{fig:ff}). 

In the feedforward SNN, the dynamics of the neurons of the first layer follows
\begin{subequations}\label{eq:simplified_SNN}
\begin{equation}
    \tau \, \frac{\rd}{\rd t} h^{(1)}_j(t) = -h^{(1)}_j(t) + \frac{\sigma}{\sqrt{p}}\sum_{\mu=1}^p \xi_{j\mu}\eta_\mu(t),
\end{equation}
and the dynamics of the neurons in the second layer follows
\begin{equation}
    \tau \, \frac{\rd}{\rd t} h^{(2)}_i(t) = -h^{(2)}_i(t) + \sum_{j=1}^{N/2}J_{ij} S^{(1)}_j(t).
\end{equation}
\end{subequations}
Analogously, in the feedforward rate network, the dynamics of the units reads
\begin{subequations}\label{eq:simplified_rate}
\begin{align}
    \tau \, \frac{\rd}{\rd t} x^{(1)}_j(t) &= -x^{(1)}_j(t) + \frac{\sigma}{\sqrt{p}}\sum_{\mu=1}^p \xi_{j\mu}\eta_\mu(t), \\
    \tau \, \frac{\rd}{\rd t}x^{(2)}_i(t) &= -x^{(2)}_i(t) + \sum_{j=1}^{N/2}J_{ij}\phi(x^{(1)}_j(t)).
\end{align}
\end{subequations}

{\color{black}In the following, we assume that the systems \eqref{eq:simplified_SNN} and \eqref{eq:simplified_rate} share the same initial conditions at time $0$, i.e. $h_j^{(1)}(0) = x_j^{(1)}(0)$ and $h_i^{(2)}(0) = x_i^{(2)}(0)$.} 

\begin{theorem}\label{theorem:4}
If $\max \phi<\infty$, for any $N>1$, and for any (feedforward) connectivity matrix $\J$, we have, for all $i$ in the second layer,
\begin{equation}
    \lim_{T\to\infty}\frac{1}{T}\int_0^T\left|h^{(2)}_i(t) - x^{(2)}_i(t)\right|\rd t \leq \sqrt{\frac{\max \phi}{2\tau}} \|\J_i\|_2. \label{eq:theorem_4}
\end{equation}
\end{theorem}
Before proving the theorem, it is useful to rewrite the dynamics of the feedforward SNN~\eqref{eq:simplified_SNN} and the feedforward rate network~\eqref{eq:simplified_rate} in a more mathematical form, which will help us to better disentangle the three independent sources of variability present in the model: (i) the quenched disorder $\bxi$, (ii) the external input signal $\{\eta_\mu\}_{\mu=1}^p$, and (iii) neuronal spike noise. Let $\{\pi_i\}_{i=1}^{N/2}$ be a collection of $N/2$ independent Poisson random measures on $\R_+\times\R_+$ with unit intensity \cite{Kin92} (see \cite{Loe17} and \cite[Ch.~1.2]{Sch22thesis} for gentle introductions on Poisson random measures in the context of spiking neuron models) and let $\{B_\mu(t)\}_{\mu=1}^p$ be $p$ independent standard Brownian motions. We can rewrite the dynamics of the feedforward SNN~\eqref{eq:simplified_SNN} as a system of stochastic differential equations 
\begin{subequations}\label{eq:simplified_SNN_SDE}
\begin{align}
    \tau \, \rd h^{(1)}_j(t) &= -h^{(1)}_j(t)\rd t + \frac{\sigma}{\sqrt{p}}\sum_{\mu=1}^p \xi_{j\mu}\rd B_\mu(t), \\
    Z^{(1)}_j(t) &= \int_{[0,t]\times\R_+}\mathbbm{1}_{z\leq \phi(h^{(1)}_j(s-))}\pi_j(\rd s,\rd z), \\
    \tau \, \rd h^{(2)}_i(t) &= -h^{(2)}_i(t)\rd t  + \sum_{j=1}^{N/2}J_{ij} \rd Z^{(1)}_j(t),
\end{align}
\end{subequations}
where $h_j(s-) = \lim_{r \to s-}h_j(r)$ denotes the left-handed limit (which implies that the dynamics are defined in terms of Itô calculus).
Formally, the spike train of neuron $j$, $S^{(1)}_j(t) = \sum_{k}\delta(t-t_j^k)$ (where the $\{t_j^k\}_k$ are the spike times of neuron $j$), is simply $S^{(1)}_j(t) = \rd Z^{(1)}_j(t)/\rd t$. The mathematical notation~\eqref{eq:simplified_SNN_SDE}, while less common in the theoretical neuroscience literature, has the important advantage of clearly separating intrinsic neuronal noise, which stems from the biophysics of single neurons and which is independent from one neuron to another, from other sources of variability. Here, intrinsic neuronal spike noise is modelled by the independent Poisson random measures $\pi_i$ \cite{Kin92,Loe17}. As stated in the main text, conditioned on the potentials $\{h^{(1)}_j(t)\}_j$, the processes $\{S^{(1)}_j(t) = \rd Z^{(1)}_j(t)/\rd t\}_j$ are independent inhomogeneous Poisson processes with instantaneous firing rates $\{\phi(h^{(1)}_j(t))\}_j$. To make the notation consistent with the mathematical notation of Itô calculus, we have replaced the formal $\{\eta_\mu(t)\rd t\}_{\mu=1}^p$ in the external input by $\{\rd B_\mu(t)\}_{\mu=1}^p$, where $\{B_\mu(t)\}_{\mu=1}^p$ are independent standard Brownian motions. Concurrently, the dynamics of the feedforward rate network~\eqref{eq:simplified_rate} becomes
\begin{subequations}\label{eq:simplified_rate_SDE}
\begin{align}
    \tau \, \rd x^{(1)}_j(t) &= -x^{(1)}_j(t)\rd t + \frac{\sigma}{\sqrt{p}}\sum_{\mu=1}^p \xi_{j\mu}\rd B_\mu(t), \label{eq:simplified_rate_SDE_1}\\
    \tau \, \frac{\rd}{\rd t}x^{(2)}_i(t) &= -x^{(2)}_i(t) + \sum_{j=1}^{N/2}J_{ij}\phi(x^{(1)}_j(t)).
\end{align}
\end{subequations}

We emphasize that in the feedforward SNN~\eqref{eq:simplified_SNN_SDE} and the feedforward rate network~\eqref{eq:simplified_rate_SDE}, the potentials of the first layers $h^{(1)}_j(t)$ and $x^{(1)}_j(t)$ are equal since they integrate the same external input $\frac{\sigma}{\sqrt{p}}\sum_{\mu=1}^p\xi_{j\mu}\rd B_{\mu}(t)$.

We can now prove Theorem~\ref{theorem:4}.
\begin{proof}
By ergodicity,
\begin{equation*}
    \lim_{T\to\infty}\frac{1}{T}\int_0^T\left|h^{(2)}_i(t) - x^{(2)}_i(t)\right|\rd t \\
    = \lim_{T\to\infty}\E_{\{\pi_j\}_{j=1}^{N/2},\{B_\mu\}_{\mu=1}^p}\left[\left|\,h^{(2)}_i(T) - x^{(2)}_i(T)\right|\right].
\end{equation*}
{\color{black}Above and in the following, expectations are taken over the Poisson random measures $\{\pi_j\}_{j=1}^{N/2}$ and Brownian motions $\{B_\mu\}_{\mu=1}^p$ but not over $\bxi$ and $\J$ which can be arbitrary.} By Jensen's inequality \cite[Theorem~5.1 p.~132]{Gut06},
\begin{equation*}
    \E_{\{\pi_j\}_{j=1}^{N/2},\{B_\mu\}_{\mu=1}^p}\left[\left|\,h^{(2)}_i(T) - x^{(2)}_i(T)\right|\right] \leq \E_{\{\pi_j\}_{j=1}^{N/2},\{B_\mu\}_{\mu=1}^p}\left[\left(h^{(2)}_i(T) - x^{(2)}_i(T)\right)^2\right]^{1/2}.
\end{equation*}
By a basic property of the conditional expectation,
\begin{equation*}
    \E_{\{\pi_j\}_{j=1}^{N/2},\{B_\mu\}_{\mu=1}^p}\left[\left(h^{(2)}_i(T) - x^{(2)}_i(T)\right)^2\right]^{1/2} = \E_{\{\pi_j\}_{j=1}^{N/2},\{B_\mu\}_{\mu=1}^p}\left[\E_{\{\pi_j\}_{j=1}^{N/2}}\left[\left(h^{(2)}_i(T) - x^{(2)}_i(T)\right)^2\Big|\{B_\mu\}_{\mu=1}^p\right]\right]^{1/2}.
\end{equation*}

It remains to study the conditional expectation:
\begin{align*}
    &\E_{\{\pi_j\}_{j=1}^{N/2}}\left[\left(h^{(2)}_i(T) - x^{(2)}_i(T)\right)^2\Big|\{B_\mu\}_{\mu=1}^p\right]\\
    &\qquad= \E_{\{\pi_j\}_{i=j}^{N/2}}\left[\left(\sum_{j=1}^{N/2}J_{ij}\int_0^T \frac{1}{\tau}e^{-(T-s)/\tau}\rd Z^{(1)}_j(s) - \sum_{j=1}^{N/2}J_{ij}\int_0^T \frac{1}{\tau}e^{-(T-s)/\tau}\phi(x^{(1)}_j(s))\rd s\right)^2\Big|\{B_\mu\}_{\mu=1}^p\right] \\
    &\qquad= \sum_{j=1}^{N/2}J^2_{ij}\,\E_{\pi_j}\left[\left(\int_0^T \frac{1}{\tau}e^{-(T-s)/\tau}\rd Z^{(1)}_j(s) - \int_0^T \frac{1}{\tau}e^{-(T-s)/\tau}\phi(x^{(1)}_j(s))\rd s\right)^2\Big|\{B_\mu\}_{\mu=1}^p\right],
\end{align*}
where in the last equality, we use the fact that the cross terms are null because the Poisson random measures $\{\pi^{(1)}_j\}_{j=1}^{N/2}$ are independent and, for all $j = 1, \dots, N/2$,
\begin{equation*}
    \E_{\pi_j}\left[\int_0^T \frac{1}{\tau}e^{-(T-s)/\tau}\rd Z^{(1)}_j(t) - \int_0^T \frac{1}{\tau}e^{-(T-s)/\tau}\phi(x^{(1)}_j(s))\rd s \;\Big|\{B_\mu\}_{\mu=1}^p\right] = 0.
\end{equation*}
On the other hand, by Itô's isometry for compensated jump processes \cite[Lemma~4.2.2 p.~197]{App09},
\begin{equation*}
    \E_{\pi_j}\left[\left(\int_0^T \frac{1}{\tau}e^{-(T-s)/\tau}\rd Z^{(1)}_j(s) - \int_0^T \frac{1}{\tau}e^{-(T-s)/\tau}\phi(x^{(1)}_j(s))\rd s\right)^2 \Big|\{B_\mu\}_{\mu=1}^p\right]
    = \int_0^T \left(\frac{1}{\tau}e^{-(T-s)/\tau}\right)^2 \phi(x^{(1)}_j(s))\rd s.
\end{equation*}
Hence, 
\begin{align*}
    \E_{\{\pi_j\}_{i=j}^{N/2}}\left[\left(h^{(2)}_i(T) - x^{(2)}_i(T)\right)^2 \;\Big|\{B_\mu\}_{\mu=1}^p\right]  &=\sum_{j=1}^{N/2}J^2_{ij}\,\int_0^T \left(\frac{1}{\tau}e^{-(T-s)/\tau}\right)^2 \underbrace{\phi(x^{(1)}_j(s))}_{\leq \max\phi}\rd s \\
    &\leq \sum_{j=1}^{N/2}J^2_{ij}\;\frac{1}{2\tau} (1 - e^{-{\color{black}2}T/\tau})\max\phi.
\end{align*}
In summary, we have
\begin{align*}
    \lim_{T\to\infty}\frac{1}{T}\int_0^T\left|h^{(2)}_i(t) - x^{(2)}_i(t)\right|\rd t
    &= \lim_{T\to\infty}\E_{\{\pi_j\}_{j=1}^{N/2},\{B_\mu\}_{\mu=1}^p}\left[\left|\,h^{(2)}_i(T) - x^{(2)}_i(T)\right|\right] \\
    &\leq \lim_{T\to\infty}\left(\sum_{j=1}^{N/2}J^2_{ij}\;\frac{1}{2\tau} (1 - e^{-{\color{black}2}T/\tau})\max\phi\right)^{1/2}\\
    &=\sqrt{\frac{\max\phi}{2\tau}}\|\J_i\|,
\end{align*}
which concludes the proof.
\end{proof}
Finally, we can verify that Theorem~\ref{theorem:weight_concentration}, i.e.
\begin{equation*}
    \|\J_{i}\|_2 \xrightarrow[N\to\infty]{\mathbb{P}} \sqrt{\frac{\alpha}{c}},
\end{equation*}
implies that as $N\to\infty$, the typical limit distance between a spiking neuron $h^{(2)}_i(t)$ and the corresponding rate unit $x^{(2)}_i(t)$, $\lim_{T\to\infty}\frac{1}{T}\int_0^T\left|h^{(2)}_{i}(t) - x^{(2)}_{i}(t)\right|\rd t$, scale as $\mathcal{O}(\sqrt{\alpha})$. Indeed, we have that 
\begin{equation*}
    \forall\varepsilon>0, \qquad \mathbb{P}\left(\lim_{T\to\infty}\frac{1}{T}\int_0^T\left|h^{(2)}_{i}(t) - x^{(2)}_{i}(t)\right|\rd t > \sqrt{\frac{\|\phi\|_\infty}{2\tau c}}\sqrt{\alpha} + \varepsilon \right) \xrightarrow[N\to\infty]{}0,
\end{equation*}
since
\begin{equation*}
    \mathbb{P}\left(\lim_{T\to\infty}\frac{1}{T}\int_0^T\left|h^{(2)}_{i}(t) - x^{(2)}_{i}(t)\right|\rd t > \sqrt{\frac{\|\phi\|_\infty}{2\tau c}}\sqrt{\alpha} + \varepsilon \right)
    \leq \mathbb{P}\left(\|\J_i\|_2 > \sqrt{\frac{\alpha}{c}} + \sqrt{\frac{2\tau}{\max\phi}}\varepsilon \right)\xrightarrow[N\to\infty]{}0.
\end{equation*}

\ensureoddstart

{\color{black}
\section{Asymmetric weights}
So far, we have studied recurrent networks with connectivity matrix
\begin{equation*}
    J_{ij} := \frac{1}{cN}\sum_{\mu=1}^p \xi_{i\mu}\left(\phi(\xi_{j\mu}) - a\right) \quad \text{for all }i\neq j,
\end{equation*}
and $J_{ii} := 0$ for all $i$. Although the connectivity matrix $\J$ is not symmetric (in the sense that $J_{ij}$ is not equal to $J_{ji}$ in general), its similarity to the connectivity of Hopfield networks may give the wrong impression that our results only apply to connectivity matrices that are somewhat symmetric. This is not the case. To show that concentration of measure in duplicate-free networks is a general phenomenon that does not require weights to be symmetric, we consider, in this section, a alternative model where weights are unambiguously asymmetric. The only difference between the original and the alternative model is that the connectivity matrix of the latter, $\J^{\mathrm{seq}}$, is inspired by that of associative networks storing sequences \cite{Ama72,Kle86,SomKan86,Her89}: 
\begin{equation}\label{eq:J_seq}
    J^{\mathrm{seq}}_{i,j} := \frac{1}{cN}\sum_{\mu=1}^p \xi_{i,\mu+1}\left(\phi(\xi_{j,\mu}) - a\right) \quad \text{for all }i\neq j,
\end{equation}
and $J_{ii} := 0$ for all $i$. By convention, $\xi_{i,p+1} = \xi_{i,1}$ for all $i$. The connectivity matrix~\eqref{eq:J_seq} can be loosely seen as that of a network storing a single cyclic sequence of $p$ patterns. As in the original model, the $N\times p$ random matrix $\bxi$ has \textit{i.i.d.}, zero-mean, unit-variance, normally distributed entries; $a:=\int_{-\infty}^\infty \mathcal{D}z\; \phi(z)$ and $c:=\int_{-\infty}^\infty \mathcal{D}z\; (\phi(z) - a)^2$, where $\mathcal{D}z$ is the standard Gaussian measure. The connectivity matrix~\eqref{eq:J_seq} is clearly asymmetric.

It is straightforward to verify that all the theoretical results of the previous sections directly apply to this alternative model with asymmetric weights. Simulations of the alternative model with asymmetric weights with $N=10^6$, load $\alpha=10^{-4}$, and the same neuronal parameters as in the main text, confirm that large SNNs can approximate the dynamics of equally large RNNs, even when the weights are unambiguously asymmetric (Fig.~\ref{fig:asym_1}). In addition, using the same alternative model and the scaling $p=N^{1/3}$ as $N\to\infty$, we confirm numerically that the dynamics of a duplicate-free sequence of SNNs with asymmetric weights can converge to the dynamics of RNNs, the average distance between the SNN and the RNN, 
\begin{equation*}
    \Delta^\mathrm{rec}_N(\alpha) := \frac{2}{N}\sum_{i=N/2+1}^{N}\lim_{T\to\infty}\frac{1}{T}\int_0^T\left|h^{\mathrm{rec}}_i(t) - x^{\mathrm{rec}}_i(t)\right|\rd t,
\end{equation*} 
lying below the theoretical bound $\sqrt{\frac{\max\phi}{2\tau c}}\sqrt{\alpha}$ (Fig.~\ref{fig:asym_2}). Note that both for the original model (Fig.~2, main text) and the alternative model (Fig.~\ref{fig:asym_2}), the theoretical bound (dashed line in both figures) is not tight; tighter bounds may be obtainable.  

In summary, this alternative model confirms that concentration of measure in duplicate-free networks also occurs in networks with asymmetric weights.

\begin{figure*}[ht]
    \includegraphics[width=\textwidth]{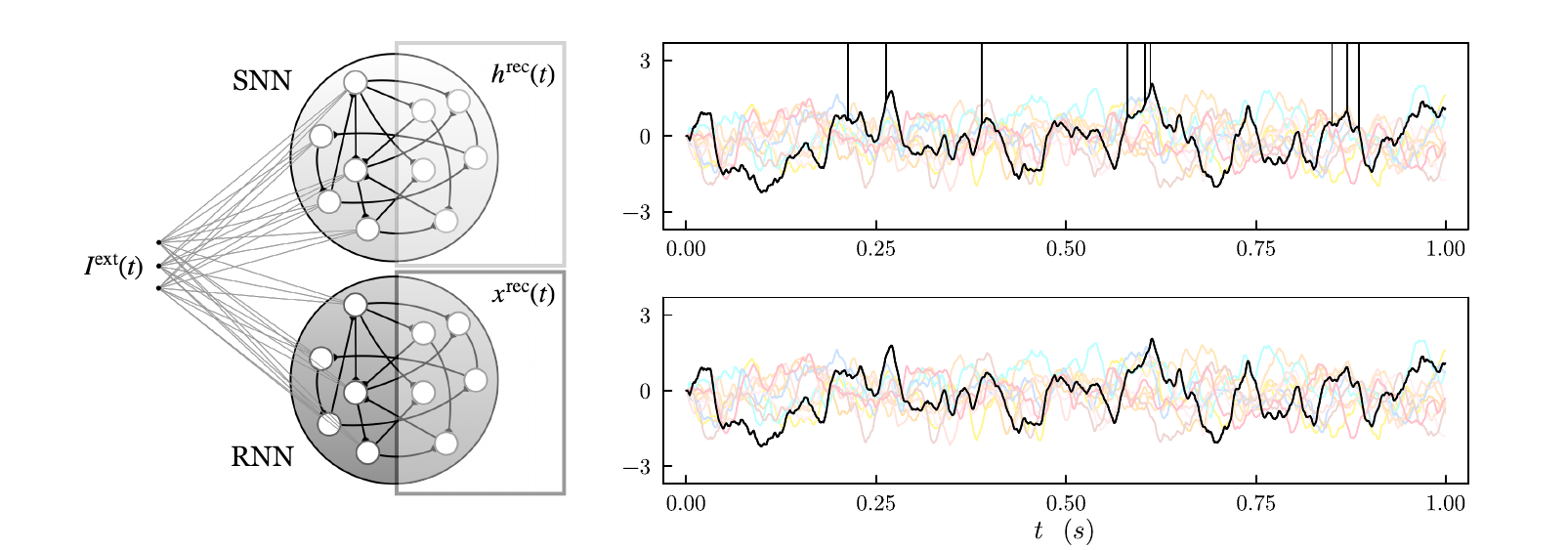}
    \caption{\label{fig:asym_1}\textbf{Large SNNs with asymmetric weights can approximate the dynamics of equally large RNNs.} Same as Fig.~1D,E in the main text except that the original connectivity matrix $\J$ is replaced by the asymmetric connectivity matrix $\J^\mathrm{seq}$ defined in~\eqref{eq:J_seq}. The networks have $N=10^6$ neurons and load $\alpha = 10^{-4}$. Same neuronal parameters as in the main text, namely, $\tau=10$~ms and $\phi(x) = \frac{1}{2\tau}\left(\tanh(x-b)+1\right)$ with $b=2$; input standard deviation is $\sigma=0.5$.}
\end{figure*}

\begin{figure*}[ht]
    \includegraphics[width=0.6\columnwidth]{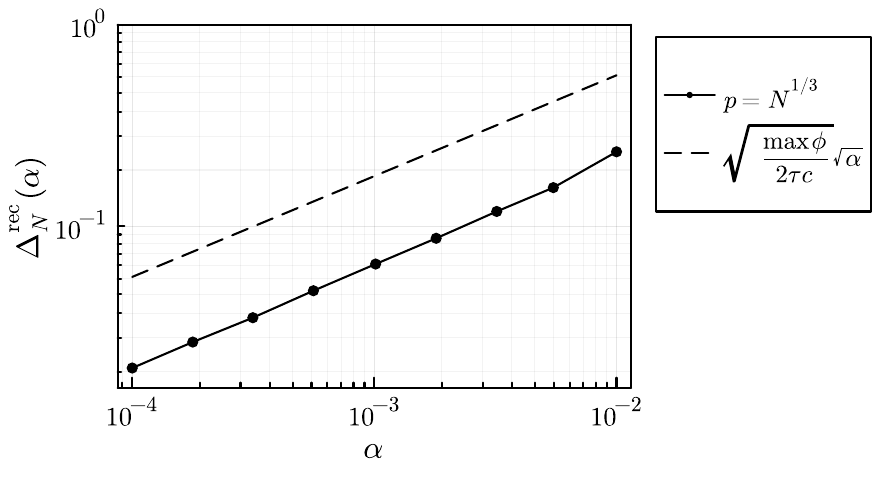}
    \caption{\label{fig:asym_2}\textbf{The dynamics of a duplicate-free sequence of SNNs with asymmetric weights can converge to the dynamics of RNNs.} Same as Fig.~2 (full circles and dashed line) in the main text except that the original connectivity matrix $\J$ is replaced by the asymmetric connectivity matrix $\J^\mathrm{seq}$ defined in~\eqref{eq:J_seq}. Simulations of $\Delta^\mathrm{rec}_N(\alpha)$ for the duplicate-free sequence of networks $p=N^{1/3}$ (full circles).
    Theoretical $\mathcal{O}(\sqrt{\alpha})$ bound predicated by the feedforward model simplification (dashed line). Same neuronal parameters as in the main text and Fig.~\ref{fig:asym_1}.}
\end{figure*}
}
\ensureoddstart
{\color{black}
\section{Adding diffusive membrane noise}

In this work, we have used networks of linear-nonlinear-Poisson neurons to give concrete examples of the concentration of measure phenomenon in networks of spiking neurons. This choice of neuron model was mainly motivated by the fact that results in this case take a particularly simple form. Namely, rate-based dynamics are described by standard RNN dynamics. However, concentration of measure does not only occur with this specific neuron model; the phenomenon also appears in networks with more general stochastic spiking neurons. To illustrate this, we briefly discuss below how Theorem~\ref{theorem:4} in Sec.~\ref{sec:feedforward} for feedforward networks of linear-nonlinear-Poisson neurons can be generalized {\color{black}to the case where diffusive membrane noise is added to the potential dynamics}.

%\subsection*{Neurons with intrinsic membrane noise}

Using the same feedfoward setup as in Sec.~\ref{sec:feedforward}, we consider the case where diffusive membrane noise is added to the membrane potential dynamics of the neurons in the first layer, i.e. the system of stochastic differential equations~\eqref{eq:simplified_SNN_SDE} is replaced by
\begin{subequations}\label{eq:SDE_membrane_noise}
\begin{align}
    \tau \, \rd h^{(1)}_j(t) &= -h^{(1)}_j(t)\rd t + \tilde{\sigma}\,\rd W_j(t)+\frac{\sigma}{\sqrt{p}}\sum_{\mu=1}^p \xi_{j\mu}\rd B_\mu(t), \label{eq:SDE_membrane_noise_1}\\
    Z^{(1)}_j(t) &= \int_{[0,t]\times\R_+}\mathbbm{1}_{z\leq \phi(h^{(1)}_j(s-))}\pi_j(\rd s,\rd z), \\
    \tau \, \rd h^{(2)}_i(t) &= -h^{(2)}_i(t)\rd t  + \sum_{j=1}^{N/2}J_{ij} \rd Z^{(1)}_j(t), \label{eq:SDE_membrane_noise_3}
\end{align}
\end{subequations}
where $W_1(t), \dots, W_{N/2}(t)$ are $N/2$ independent standard Brownian motions that are also independent from the external input {\color{black}signals} $B_1(t), \dots, B_{p}(t)$, and $\tilde{\sigma}>0$ is the intrinsic membrane noise parameter. The diffusive noise $\tilde{\sigma}\,\rd W_j(t)$ in~\eqref{eq:SDE_membrane_noise_1} should be interpreted as intrinsic membrane noise due, for example, to the effect of channel noise on membrane potential dynamics. Note that, as in the original feedforward setup in Sec.~\ref{sec:feedforward}, neurons in the second layer, Eq.~\eqref{eq:SDE_membrane_noise_3}, simply follow the dynamics of passive membranes {\color{black}(without membrane noise)}. 

{\color{black}We recall that, in Sec.~\ref{sec:feedforward}, Theorem~\ref{theorem:4} gives a bound for the deviation of the potentials $h^{(2)}_i(t)$ in the second layer from their expectations $x^{(2)}_i(t) = \E_{\{\pi_j\}_{j=1}^{N/2}}\left[h^{(2)}_i(t)\,|\,\{B_\mu\}_{\mu=1}^p\right]$ conditioned on the external input signals $\{B_\mu(t)\}_{\mu=1}^p$. In order to obtain an analog of Theorem~\ref{theorem:4} for the system with membrane noise~\eqref{eq:SDE_membrane_noise}, our first task is to derive equations for the expectations
\begin{equation*}
    x^{(2)}_i(t) = \E_{\{W_j,\pi_j\}_{j=1}^{N/2}}\left[h^{(2)}_i(t)\,|\,\{B_\mu\}_{\mu=1}^p\right]
\end{equation*}
corresponding to the system~\eqref{eq:SDE_membrane_noise}. By the linearity of the expectation, we have
\begin{equation*}
    \tau \, \frac{\rd}{\rd t}x^{(2)}_i(t) = -x^{(2)}_i(t) + \sum_{j=1}^{N/2}J_{ij}\,\E_{\{W_j,\pi_j\}_{j=1}^{N/2}}\left[\phi(h^{(1)}_j(t)) \big| \{B_\mu\}_{\mu=1}^p\right].
\end{equation*}
Thus, to derive a closed system of equations for the $x^{(2)}_i(t)$, it suffices to notice that the expected firing rates $r^{(1)}_j(t):=\E_{\{W_j,\pi_j\}_{j=1}^{N/2}}\left[\phi(h^{(1)}_j(t)) \big| \{B_\mu\}_{\mu=1}^p\right]$ of each individual neuron $j=1, \dots, N/2$ in the first layer are given by the solutions of Fokker-Planck equations (conditioned on the external inputs $I^{\mathrm{ext}}_j(t)\rd t = \frac{\sigma}{\sqrt{p}}\sum_{\mu=1}^p \xi_{j\mu}\rd B_\mu(t)$) describing the evolution of a time-varying probability densities $q^{(1)}_j(\cdot,t)$:
\begin{align*}
    r^{(1)}_j(t) &= \int_\R \phi(x)q^{(1)}_j(x,t)\rd x\\
    \tau\,\frac{\partial}{\partial t} q^{(1)}_j(x,t) &= \frac{\partial}{\partial x}\left([x- I^{\mathrm{ext}}_j(t)]q^{(1)}_j(x,t)\right) + \frac{\tilde{\sigma}^2}{2\tau}\frac{\partial^2}{\partial x^2}q^{(1)}_j(x,t),
\end{align*}
with initial conditions $q^{(1)}_j(x,0) = \delta(x - h^{(1)}_j(0))$ ($\delta$ denotes the Dirac delta function). The Fokker-Planck equation above has an explicit solution 
\begin{equation*}
    q^{(1)}_j(x,t) = \frac{1}{\sqrt{2 \pi \Sigma(t)}}\exp\Bigg(-\frac{(x-\bar{x}^{(1)}_j(t))^2}{2\Sigma(t)}\Bigg),
\end{equation*}
where the mean potential $\bar{x}^{(1)}_j(t)$ follows the same dynamics as \eqref{eq:simplified_rate_SDE_1}, i.e.
\begin{equation*}
    \tau \, \rd \bar{x}^{(1)}_j(t) = -\bar{x}^{(1)}_j(t)\rd t + \frac{\sigma}{\sqrt{p}}\sum_{\mu=1}^p \xi_{j\mu}\rd B_\mu(t),
\end{equation*}
and the time-varying variance $\Sigma(t)$ is
\begin{equation*}
    \Sigma(t) = \frac{\tilde{\sigma}^2}{2\tau}(1 - e^{-2t/\tau}).
\end{equation*}
This gives us generalized rate-based equations for the system with membrane noise~\eqref{eq:SDE_membrane_noise}:
\begin{subequations}\label{eq:rate_membrane_noise}
\begin{align}
    \tau \, \rd \bar{x}^{(1)}_j(t) &= -\bar{x}^{(1)}_j(t)\rd t + \frac{\sigma}{\sqrt{p}}\sum_{\mu=1}^p \xi_{j\mu}\rd B_\mu(t) \\
    \tau \, \frac{\rd}{\rd t}x^{(2)}_i(t) &= -x^{(2)}_i(t) + \sum_{j=1}^{N/2}J_{ij}\Phi(\bar{x}^{(1)}_j(t), \Sigma(t)),
\end{align}
\end{subequations}
where the variance-dependent transfer function $\Phi(\bar{x},\Sigma)$ is
\begin{equation*}
    \Phi(\bar{x}, \Sigma) = \int_{\R}\phi(x)\frac{1}{\sqrt{2 \pi \Sigma}}\exp\Bigg(-\frac{(x-\bar{x})^2}{2\Sigma}\Bigg)\rd x.
\end{equation*}
%Since Eqs.~\eqref{eq:FP_membrane_noise} do not form a simple system of $N$ ordinary differential equations, they are not standard rate-based equations. In particular, the first layer potentials $x^{(1)}_j(t)$ in~\eqref{eq:simplified_rate_SDE} have been replaced by the time-varying probability densities $q^{(1)}_j(\cdot,t)$. However, the system~\eqref{eq:FP_membrane_noise} still describes rate-based dynamics in the following sense: the potentials in the second layer $x^{(2)}_j(t)$ behave \textit{as if} neurons in the first layer were transmitting their expected firing rates
%\begin{equation*}
%    r^{(1)}_j(t) = \E_{\{W_j,\pi_j\}_{j=1}^{N/2}}\left[\phi(h^{(1)}_j(t)) \big| \{B_\mu\}_{\mu=1}^p\right] = \int_\R \phi(x)q^{(1)}_j(x,t)\rd x.
%\end{equation*}

Assuming that the transfer function $\phi$ is bounded, i.e. $\max\phi< \infty$, the statement and the proof of Theorem~\ref{theorem:4} can be readily adapted to the system~\eqref{eq:SDE_membrane_noise} if the original rate-based system~\eqref{eq:simplified_rate_SDE} is replaced by the generalized rate-based dynamics~\eqref{eq:rate_membrane_noise}. 

%We stress that the final result will \textit{not} imply that the system with membrane noise~\eqref{eq:SDE_membrane_noise} can be well approximated by the original rate-based ordinary differential equations~\eqref{eq:simplified_rate_SDE}.
}

%In~\eqref{eq:FP_membrane_noise}, the deterministic potentials $x^{(1)}_j(t)$ of \eqref{eq:simplified_rate_SDE} have been replaced by the time-varying probability densities $q^{(1)}_j(\cdot, t)$ and the deterministic firing rates $\phi(x^{(1)}_j(t)) = \phi(h^{(1)}_j(t))$ of \eqref{eq:simplified_rate_SDE} have been replaced by the expected firing rates
%\begin{equation*}
%    \int_\R \phi(x)q^{(1)}_j(x,t)\rd x = \E\left[\phi(h^{(1)}_j(t)) \big| \{B_\mu\}_{\mu=1}^p\right].
%\end{equation*}
%Importantly, the system~\eqref{eq:FP_membrane_noise} does describe rate-based dynamics since units in the second layer receive the expected firing rates $\int_\R \phi(x)q^{(1)}_j(x,t)\rd x$ of the units in the first layer.

To adapt the proof of Theorem~\ref{theorem:4}, we simply need to notice that
\begin{multline*}
    \E_{{\color{black}W_j,\pi_j}}\left[\left(\int_0^T \frac{1}{\tau}e^{-(T-s)/\tau}\rd Z^{(1)}_j(s) - \int_0^T \frac{1}{\tau}e^{-(T-s)/\tau}\int_\R \phi(x)q^{(1)}_j(x,s)\rd x\,\rd s\right)^2 \Big|\{B_\mu\}_{\mu=1}^p\right] \\
    = \Var_{{\color{black}W_j,\pi_j}}\left(\int_0^T \frac{1}{\tau}e^{-(T-s)/\tau}\rd Z^{(1)}_j(s)\,\Big|\,\{B_\mu\}_{\mu=1}^p\right),
\end{multline*}
and by the law of total variance,
\begin{multline*}
    \Var_{{\color{black}W_j,\pi_j}}\left(\int_0^T \frac{1}{\tau}e^{-(T-s)/\tau}\rd Z^{(1)}_j(s)\,\Big|\,\{B_\mu\}_{\mu=1}^p\right) \\
    = \E_{W_j}\left[\Var_{\pi_j}\left(\int_0^T \frac{1}{\tau}e^{-(T-s)/\tau}\rd Z^{(1)}_j(s)\,\Big|\,W_j,\{B_\mu\}_{\mu=1}^p\right)\right] \\
    + \Var_{{\color{black}W_j}}\left(\E_{\pi_j}\left[\int_0^T \frac{1}{\tau}e^{-(T-s)/\tau}\rd Z^{(1)}_j(s)\,\Big|\,W_j,\{B_\mu\}_{\mu=1}^p\right]\right).
\end{multline*}
But since
\begin{align*}
    \Var_{\pi_j}\left(\int_0^T \frac{1}{\tau}e^{-(T-s)/\tau}\rd Z^{(1)}_j(s)\,\Big|\,W_j,\{B_\mu\}_{\mu=1}^p\right) &= \int_0^T \left(\frac{1}{\tau}e^{-(T-s)/\tau}\right)^2\int_\R \phi(h^{(1)}_j(s))\,\rd s,\\
    \E_{\pi_j}\left[\int_0^T \frac{1}{\tau}e^{-(T-s)/\tau}\rd Z^{(1)}_j(s)\,\Big|\,W_j,\{B_\mu\}_{\mu=1}^p\right] &= \int_0^T \frac{1}{\tau}e^{-(T-s)/\tau} \phi(h^{(1)}_j(s))\rd s,
\end{align*}
we obtain
\begin{align}\label{eq:LTV}
    &\Var_{{\color{black}W_j,\pi_j}}\left(\int_0^T \frac{1}{\tau}e^{-(T-s)/\tau}\rd Z^{(1)}_j(s)\,\Big|\,\{B_\mu\}_{\mu=1}^p\right) \nonumber\\
    &\qquad= \int_0^T \left(\frac{1}{\tau}e^{-(T-s)/\tau}\right)^2\int_\R \phi(x)q^{(1)}_j(x,s)\rd x\,\rd s 
    + \Var_{{\color{black}W_j}}\left(\int_0^T \frac{1}{\tau}e^{-(T-s)/\tau} \phi(h^{(1)}_j(s))\rd s\,\Big|\,\{B_\mu\}_{\mu=1}^p\right).
\end{align}
The second term on the right-hand side of~\eqref{eq:LTV} implies that the final bound for the network with membrane noise~\eqref{eq:SDE_membrane_noise} has to be looser than the bound~\eqref{eq:theorem_4} for the network with linear-nonlinear-Poisson neurons.
From~\eqref{eq:LTV}, we can easily derive the final bound
\begin{equation*}
    \lim_{T\to\infty}\frac{1}{T}\int_0^T\left|h^{(2)}_i(t) - x^{(2)}_i(t)\right|\rd t \leq \left(\sqrt{\frac{\max \phi}{2\tau}} + \max\phi\right) \|\J_i\|_2.
\end{equation*}
Note that this bound might not be optimal and {\color{black}a} tighter bound may be {\color{black}obtainable}.}

{\color{black}The model treated in this section is instructive as it illustrates an interesting fact: the potentials in the second layer $h^{(2)}_j(t)$ can converge to $x^{(2)}_j(t)$ in large networks (if the norms $\|\J_i\|_2$ converge to $0$) even if the the potentials in the first layer $h^{(1)}_j(t)$ do not converge to the mean potentials $\bar{x}^{(1)}_j(t)$. Indeed, in this model, we know that the variance of $h^{(1)}_j(t) - \bar{x}^{(1)}_j(t)$ is $\Sigma(t)$, which does not depend on the network size. This shows that concentration of measure can occur even in cases where membrane potentials dynamics are intrinsically noisy.}

{\color{black}We mention that whether or not the concentration of measure phenomenon allows for the exact reduction of the dynamics of large SNNs to rate-based dynamics described by systems of \textit{ordinary} differential equations (as in a RNN) depends on the spiking neuron model considered. We have shown that such a reduction is possible for Linear-Nonlinear-Poisson neurons and Linear-Nonlinear-Poisson neurons with diffusive membrane noise, but this is not true in general. The derivation of the generalized equations describing the exact rate-based dynamics in duplicate-free networks of spiking neurons including the effects of refractoriness \cite{Ger95}, adaptation \cite{PozNau13}, and short-term synaptic plasticity \cite{Sch22} is an outstanding problem that is beyond the scope of this work.}

\bibliography{supplement}